\documentclass{aa}  

\usepackage{natbib}
\bibpunct{(}{)}{;}{a}{}{,}
\usepackage{graphicx}
\usepackage{float}
\usepackage{subcaption}
\usepackage[normalem]{ulem}
\usepackage{txfonts}
\usepackage{amsmath}
\usepackage{svg}
\usepackage{longtable}
\usepackage{multicol}
\usepackage{afterpage}
\usepackage{placeins}
\usepackage{bm}
\usepackage{dblfloatfix}
\usepackage[all]{nowidow}
\usepackage{longtable} 
\usepackage{pdflscape} 
\usepackage{array}
\usepackage{adjustbox}
\usepackage[colorlinks=true, linkcolor=black, urlcolor=blue, citecolor=blue]{hyperref}

\makeatletter
\newcommand{\showdimensions}{
  Font size: \number\f@size~pt\\
  Column width: \the\columnwidth\\
  Full text width: \the\textwidth
}
\makeatother

\begin{document} 

   \title{Extending TESS flare frequency distributions with CHEOPS: \\ Power-law versus lognormal}

   \subtitle{}

   \author{J. Poyatos
          \inst{1,2} \href{https://orcid.org/0000-0003-4064-4268}{\protect\includegraphics[height=0.25cm]{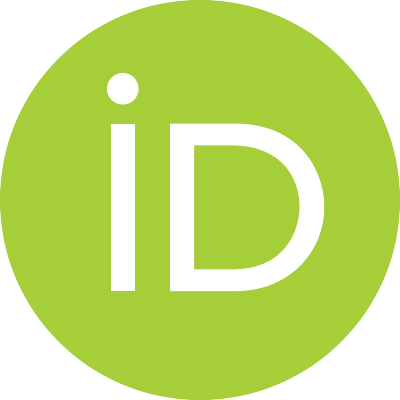}}
          \and
          O. Fors\inst{1,2} \href{https://orcid.org/0000-0002-4227-9308}{\protect\includegraphics[height=0.25cm]{Fig/orcid.pdf}}
          \and
          J.M. Gómez Cama\inst{2,3} \href{https://orcid.org/0000-0003-0173-5888}{\protect\includegraphics[height=0.25cm]{Fig/orcid.pdf}}
          }

   \institute{Departament de Física Quàntica i Astrofísica, Institut de Ciències del Cosmos (ICCUB), Universitat de Barcelona (IEEC-UB),
Martí i Franquès 1, E-08028 Barcelona, Spain\\
\email{julienpoyatos@icc.ub.edu}
        \and
            Institut d'Estudis Espacials de Catalunya (IEEC), 08860 Castelldefels (Barcelona), Spain
         \and
             Departament d’Enginyeria Electrònica i Biom{\`e}dica, Institut de Ciències del Cosmos (ICCUB), Universitat de Barcelona, IEEC-UB, Martí i Franquès 1,
E-08028 Barcelona, Spain
             }

   \date{Received 16 October 2025; accepted 11 March 2026}

 \abstract
   {Stellar flares are intense bursts of radiation caused by magnetic reconnection events on active stars. They are especially frequent on M dwarfs, where they can significantly influence the habitability of orbiting planets. Flare frequency distributions (FFDs) are typically modelled as power laws. However, recent studies challenge this assumption and propose alternative distributions such as lognormal laws that imply different flare generation mechanisms and levels of planetary impact.}
   {This study investigates which statistical distribution best describes flare occurrences on M dwarfs, considering both equivalent duration (ED), the quantity directly measured from light-curve photometry, and bolometric energy, which is relevant for physical interpretation and habitability assessment.}
   {We analysed 110 M dwarfs observed with TESS and CHEOPS, detecting 5620 flares. We decomposed complex flare events, corrected for detection biases in the recovery rate and energy estimation, and scaled the FFDs from both missions to construct a combined distribution covering six orders of magnitude in bolometric energy.}
   {We find that ED-based FFDs closely follow a power-law distribution, reflecting the intrinsic photometric flare occurrence. However, bolometric-energy-based FFDs deviate significantly from a pure power law. They are better described by a lognormal distribution, although the best fit is achieved with a truncated power law, exhibiting a break at $1.8 \times 10^{35}$ erg. Using right-tail-stabilised Kolmogorov-Smirnov and exceedance tests, we attribute this deviation to limited sampling of the most energetic events.}
   {Our results show that the low-energy flattening, previously interpreted as evidence of lognormal behaviour, arises from observational biases and can be corrected through flare injection-recovery and the combination of observations from instruments with different sensitivities. We also find that current instruments are unable to reliably sample flares above $10^{35}$ erg, which are the most relevant for exoplanetary atmospheric effects. The upcoming PLATO mission will be able to investigate both regimes.}

   \maketitle

%\showdimensions

\section{Introduction}
Stellar flares are intense bursts of energy caused by magnetic reconnection in a star's atmosphere, which release large amounts of radiation across the electromagnetic spectrum. Magnetic activity is especially pronounced on M dwarfs, making flares more frequent and energetic relative to their quiescent luminosity than on solar-type stars. This increased activity has important implications for exoplanet habitability: energetic flares can erode planetary atmospheres and expose planetary surfaces to harmful ultraviolet and X-ray radiation \citep{Segura_2010, Howard_2018, Tilley_2019}. Conversely, stellar flares may be the only source of increased ultraviolet radiation needed to induce prebiotic photochemical reactions on planets orbiting M dwarfs \citep{Rimmer_2018}. As M dwarfs are the most common hosts of terrestrial exoplanets in the habitable zone, understanding the flare activity on these stars is crucial to assessing their potential to support life \citep{Ribas_2023}. 

To characterise flare activity, large photometric surveys such as the Transiting Exoplanet Survey Satellite (TESS) have become central. TESS offers long-duration, near-continuous observations of large samples of stars up to a 20-second cadence, which is ideal for detecting stochastic events such as flares and constructing flare frequency distributions (FFDs) \citep{Gunther_2020, Pietras_2022}. In contrast, the CHaracterising ExOPlanet Satellite (CHEOPS) observes a smaller number of targets with ultra-high photometric precision and shorter observational windows (typically 99 minutes per visit) but at higher cadence (up to 3 seconds) \citep{Benz_2021}. This complementary observing strategy enables CHEOPS to probe short-duration, low-energy flares that TESS cannot recover, extending the lower-energy end of the FFDs \citep{Bruno_2024, Poyatos_2025}. Together, the two missions offer a unique opportunity to characterise the full energy range of flare activity on M dwarfs. 

A significant fraction of flares exhibit complex structures with multiple peaks or sub-events. High-cadence observations from TESS and CHEOPS are particularly valuable for resolving these substructures, which are often missed at lower cadence \citep{Howard_2022}. Two physical interpretations have been proposed to explain complex flare substructures \citep{Davenport_2014}. In the first scenario, the multiple flare components within a complex event are physically associated, occurring within a single active region \citep{Kowalski_2010, Anfinogentov_2013}. In the second scenario, complex flares result from the superposition of independent flares originating from nearby active regions, often referred to as sympathetic flares \citep{Moon_2002, Bruno_2024}. Understanding complex flares is crucial because treating them as single events or decomposing them into components can alter derived FFDs.

Flare frequency distributions are commonly modelled using a power-law relationship between flare frequency and energy, expressed as $dN / dE \propto E^{-\alpha}$, motivated by both theoretical considerations and solar observations \citep{Longscope_2000, Aschwanden_2000}. This approach enables convenient comparison of $\alpha$ indexes between samples and the extrapolation of flare occurrence rates beyond the observed regime, which is essential for estimating the cumulative impact of flares on exoplanet atmospheres over long timescales \citep{Gunther_2020, Feinstein_2024}. However, several recent studies suggest that the power-law assumption must first be justified and that it may break down at low and high energy if the studied range is broad enough \citep{Ryan_2016, Verbeeck_2019, Sakurai_2022}. These deviations may arise from a combination of observational biases, such as reduced detection efficiency at low amplitudes, extreme-value statistics at high energies, and intrinsic physical processes, such as a turnover or truncation in the flare energy distribution \citep{Howard_2019, Gao_2022}. 

In particular, \cite{Bruno_2024} and \cite{Poyatos_2025} reported that when complex flares are decomposed into individual components, M dwarf FFDs tend to be better described by a lognormal distribution. This alternative description suggests a different flare generation scenario, in which strong flares may result from the multiplicative superposition of smaller events triggering reconnection cascades in neighbouring regions, often referred to as sympathetic flares, rather than emerging from a purely self-organised criticality process \citep{Aschwanden_2021}. Supporting this view, \cite{Howard_2022} showed, using TESS 20- and 120-sec observations of low-mass flare stars, that 42\% of flares with energies higher than $10^{33}$ erg exhibited complex substructures, increasing to 70\% for flares above $10^{35}$ erg. This suggests that energetic flares more often have complex substructures, likely caused by the superposition of smaller events. Considering each of these sub-events as an individual flare would therefore affect the resulting FFD shape, effectively reducing the frequency of high-energy flares while increasing that of low-energy ones. 

When constructing FFDs, the choice of energy metric can significantly affect the observed distribution. The FFDs based on equivalent duration (ED), which is measured directly from photometry, generally follow a power law, reflecting predictable scaling with flare energy. In contrast, FFDs based on bolometric energy ($E_{\text{bol}}$) are obtained by multiplying the ED by the stellar quiescent luminosity. For broad stellar samples, converting the ED into bolometric energy can produce distributions that deviate from a simple power law due to the influence of the stellar luminosity distribution. Since stellar luminosities across large samples are approximately lognormally distributed, the convolution of a power-law ED distribution with this luminosity distribution can distort the resulting energy-based FFD, as observed in large flare samples that deviate from a pure power law \citep{Howard_2019, Pietras_2022}. This effect complicates comparisons of FFDs across different stellar samples. It can make the distribution appear to deviate from a pure power law even if the underlying flare generation process is unchanged. Understanding this convolution is therefore essential for accurately interpreting energy-based FFDs and identifying genuine deviations from canonical flare scaling laws. 

In this work, we revisit the statistical shape of FFDs on a sample of 110 M dwarfs by combining flare observations from both TESS and CHEOPS. We first describe our flare detection and characterisation using the \texttt{AltaiPony} package, decompose complex flare structures using a method similar to \cite{Davenport_2014}, and apply consistent recovery and energy corrections to both datasets to address observational biases following \cite{Gao_2022}. We then construct and fit FFDs based on ED and bolometric energy, comparing the performance of power-law, lognormal, and truncated power-law models using likelihood-based statistical tests. Finally, we discuss the implications of the observed deviations from the canonical power law, particularly in the context of flare physics and the long-term habitability of exoplanets orbiting M dwarfs. 

The structure of the paper is as follows: in Section \ref{section:methods}, we describe the data sources, flare detection, decomposition, and correction methods, energy calibration, and the construction of the FFDs. In Section \ref{section:results}, we present the fitted distributions for each dataset and energy metric. In Section \ref{section:discussion}, we examine our findings in the broader context of stellar flare physics and discuss their implications for future studies, especially in the context of the PLATO mission. Finally, we summarise our conclusions in Section \ref{section:conclusion}.  
 
\section{Methods}
\label{section:methods}

\begin{figure*}
\centering
\includegraphics[width=0.85\textwidth]{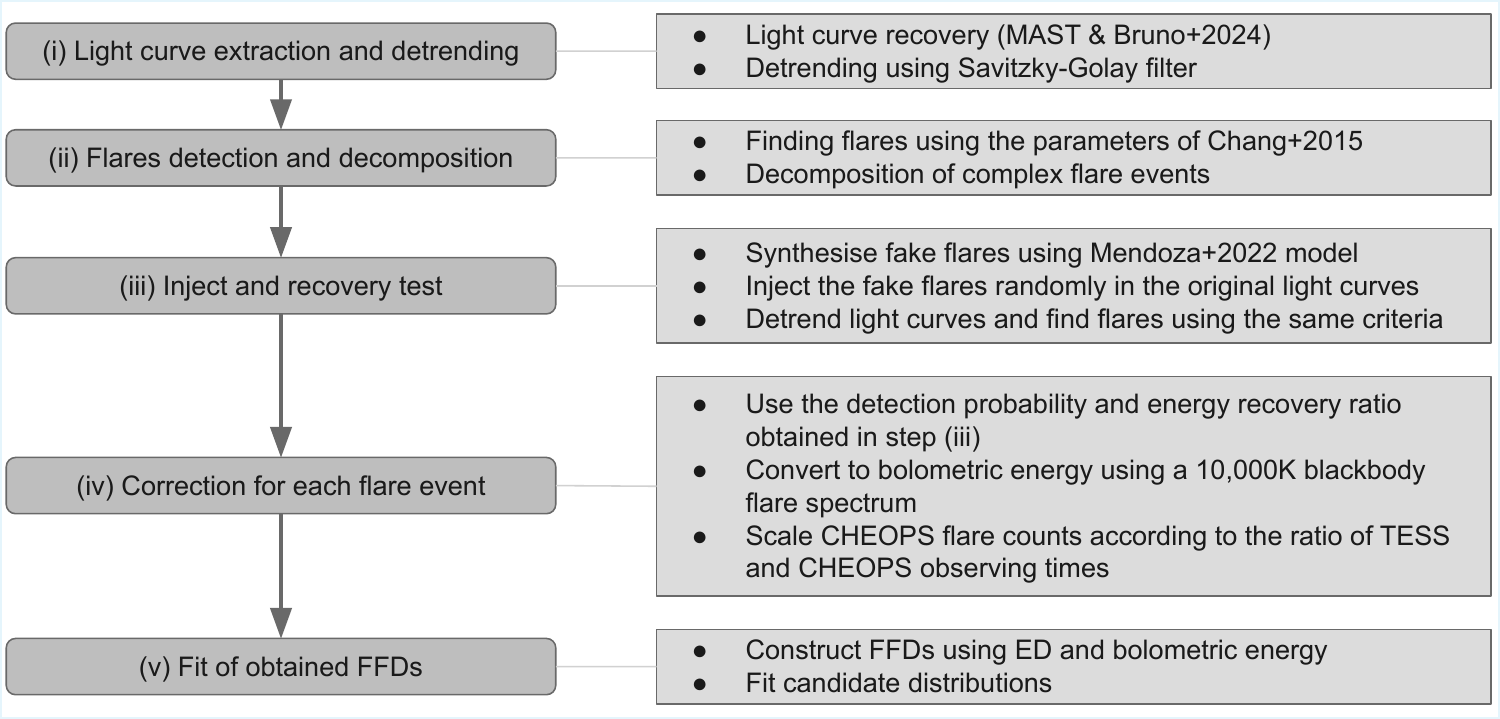}
\caption{Flow diagram of the analysis procedure used in this study. The left bubbles describe the main steps, while the right squares provide additional details.}
\label{fig:procedure}
\end{figure*}

This section describes the full analysis procedure, from sample construction to the generation and fitting of FFDs. An overview of the procedure is shown in Figure \ref{fig:procedure}. We begin by describing the target selection and stellar parameters, then detail the detrending and flare detection routines applied to the CHEOPS and TESS light curves. We further present the flare energy calibration and injection-recovery tests used to correct for detection biases, followed by the statistical methods employed to fit candidate distributions to the FFDs. Finally, we describe the merging of CHEOPS and TESS samples into a combined FFD that extends the accessible energy range.

\subsection{Target selection}
Several CHEOPS programmes have targeted late-type stars. To construct our sample, we selected all M dwarfs observed by the completed programmes available in the CHEOPS archive. In particular, we combined data from two programmes: CH\_PR100018 (PI: I. Pagano), a guaranteed time observation (GTO) programme dedicated to characterising the variability of late-K and M dwarfs since April 2020, and CH\_PR100010 (PI: G. Szabó), which focused on monitoring the debris disc of AU Mic between July 2020 and December 2023. These programmes collectively yield around 52 days of CHEOPS observations across 110 M0 to M5 dwarfs. Therefore, the stellar sample used in this work is the same as in \cite{Poyatos_2025}.

\begin{figure}
\centering
\includegraphics[width=0.9\columnwidth]{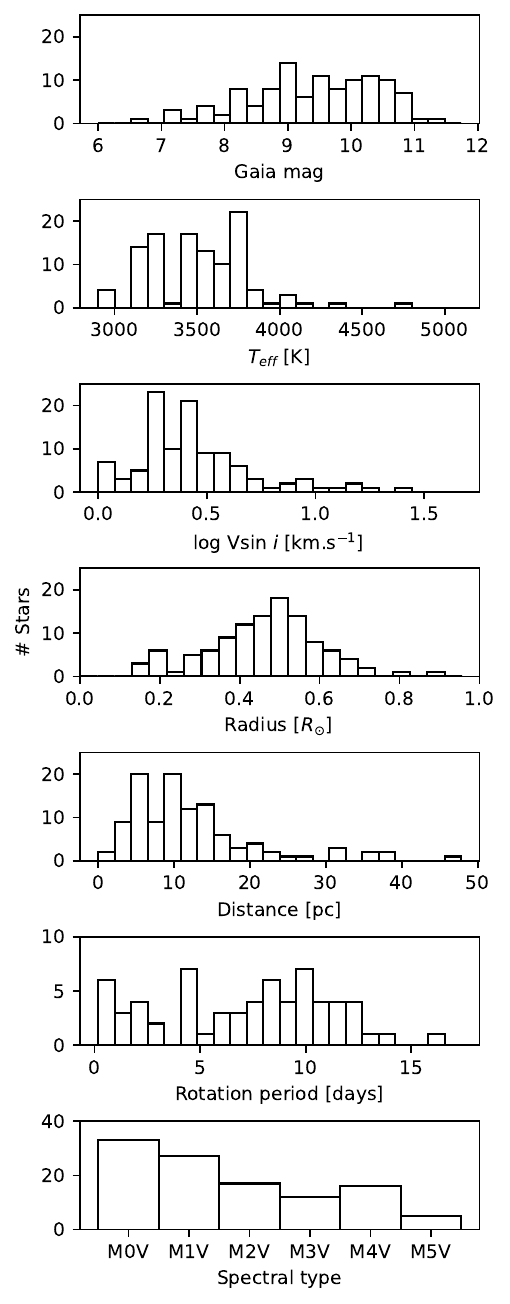}
\caption{Parameters of the stellar sample. Panels from top to bottom show histograms of Gaia $G$-band magnitude, $T_{eff}$, log(V sin i), radius, distance, rotation period, and spectral type.}
\label{fig:star_description}
\end{figure}

We obtained stellar parameters from external catalogues: Gaia DR3 \citep{Gaia_2023} provided $G$-band magnitudes, effective temperatures, and distances derived from parallaxes, while stellar radii were taken from the TESS Input Catalogue \citep{TESS_2018}. Spectral types were retrieved from the CHEOPS observation metadata. We also searched for chromospheric activity indicators, including $H_\alpha$ equivalent widths from \cite{Carmenes_2019} and $\log R'_{HK}$ values from \cite{HARPS_2017} and \cite{Boro_2018}. However, these indicators were available for only 64 and 71 of the 110 targets, respectively, limiting their usefulness for a comprehensive statistical comparison. As an alternative proxy for activity, we adopted the average projected rotational velocities (V sin i) of \cite{Bruno_2024}. Rotation periods were determined as the dominant periodicities in the TESS light curves using a Lomb-Scargle periodogram. Rotation periods longer than 27 days could not be measured due to the typical TESS sector observation duration. Figure \ref{fig:star_description} shows the distribution of these parameters, while Table \ref{annex:table}\footnote{Table \ref{annex:table} is available in electronic form at the CDS via anonymous ftp to \href{cdsarc.u-strasbg.fr}{cdsarc.u-strasbg.fr}(130.79.128.5) or via \href{http://cdsweb.u-strasbg.fr/cgi-bin/qcat?J/A+A/}{http://cdsweb.u-strasbg.fr/cgi-bin/qcat?J/A+A/}.} lists their values for each star.

\begin{figure}
\centering
\includegraphics[width=0.9\columnwidth]{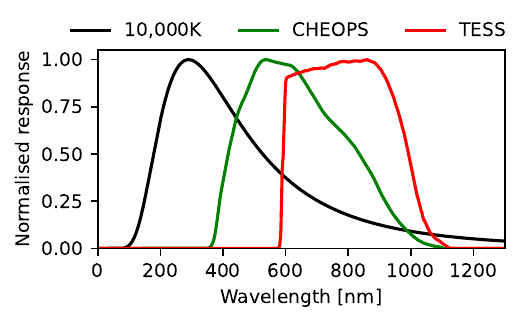}
\caption{Response function of CHEOPS (green) and TESS (red), shown together with a 10,000 K flare blackbody spectrum (black).}
\label{fig:blackbody}
\end{figure}

In Figure \ref{fig:blackbody}, we compare the CHEOPS and TESS spectral response functions with a 10,000K blackbody spectrum, a typical approximation for M dwarf flare spectra \citep{Kowalski_2024}. The TESS response is redder than CHEOPS's, making it less sensitive to the blue-weighted emission of flares \citep{Berger_2024}. Combined with the faster cadence and superior photometric precision of CHEOPS, this implies a higher sensitivity to weaker and shorter-duration flares. Conversely, the longer observing times and wide field of TESS favour detecting rare, high-energy flares. 

\begin{figure}
\centering
\includegraphics[width=0.9\columnwidth]{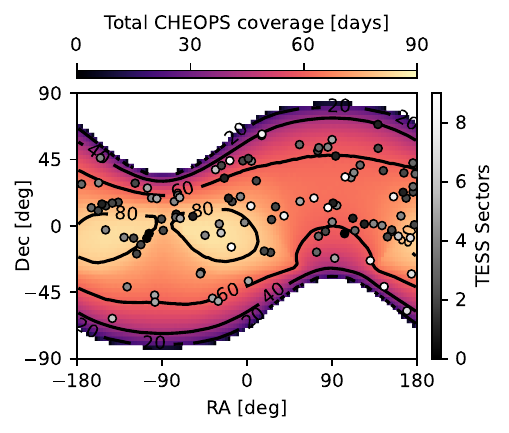}
\caption{Coordinates of the target stars overlaid on the CHEOPS sky coverage map, with a minimum observing duration of 39 minutes per orbit, corresponding to $\sim$40\% efficiency. Circle colours show the number of TESS sectors covering each target. Background shading indicates total CHEOPS observing time in days per year, with black contours marking 20, 40, 60, and 80 days.}
\label{fig:map}
\end{figure}

Figure \ref{fig:map} illustrates the CHEOPS yearly field-of-view coverage overlaid with the target positions. Even though CHEOPS's optimal coverage lies near the ecliptic plane, and TESS observes with high frequency around the ecliptic poles, these missions enable combined flare monitoring across a broad portion of the sky.

\subsection{Light curve detrending}
\label{subsection:detrending}
The combination of high photometric precision, short cadence, and extensive sky coverage makes TESS a powerful mission to study flares at our targets. We downloaded all available TESS light curves using the \texttt{lightkurve} package, selecting both 'short' (2-minute) and 'fast' (20-second) cadence data from the Mikulski Archive for Space Telescopes. This resulted in 454 light curves across 107 targets, spanning observations up to sector 93, for a cumulative total of around 8,670 days.

We used the Pre-search Data Conditioning Simple Aperture Photometry (PDCSAP) flux, which has instrumental systematics corrected by the SPOC pipeline \citep{Jenkins_2016}. Although PDCSAP processing may introduce spurious effects (e.g. \citet{Nardiello_2022}), the timescales of flares and instrumental trends are sufficiently distinct in our data to justify its use over the uncorrected SAP flux. 

To remove rotational modulation, we applied a third-order Savitzky-Golay filter. Following \cite{Davenport_2016} and \cite{Yang_2018}, we selected the filtering window as one fourth of the dominant period identified by a Lomb-Scargle periodogram. Periods were accepted if the power exceeded 0.1 and the false alarm probability was below 0.01. When no significant period was identified, one fourth of the full length of the light curve was adopted. Filtering was performed iteratively, masking data points that deviate more than $3\sigma$ above the fit to mitigate flare contamination during detrending. 

CHEOPS light curves were obtained from \cite{Bruno_2024}, who performed photometry on 60-pixel imagettes centred on the target stars. Imagettes photometry enables CHEOPS's highest cadence of 3 seconds. Individual visits were combined to create a single light curve per target. In total, we obtained around 52 days of observation for the 110 M dwarfs. The light curves were processed with the CHEOPS Data Reduction Tool \citep{Morgado_2022, Fortier_2024} and normalised using low-order polynomial fits. Full details of imagettes processing can be found in \cite{Bruno_2024}.
The same Savitzky-Golay filter was applied to the CHEOPS data, using the rotation period identified from TESS where available. If no period was found, or if the target had no TESS light curve, we used one fourth of the CHEOPS light curve duration as the filtering window.

\subsection{Flare detection and decomposition}
\label{subsection:detection}
Flares were identified using the \texttt{AltaiPony} package \citep{Ilin_2021_altaipony}, which applies the criteria outlined in Equation 3 of \cite{Chang_2015}. To be detected, a flare must:
\begin{itemize}
    \item consist of positive deviations from the quiescent flux level,
    \item exceed $N_1\sigma$ above the local scatter,
    \item exceed $(N_2\sigma + \text{photometric error})$ above the local scatter,
    \item and consists of at least $N_3$ consecutive points satisfying these conditions.
\end{itemize}
We adopted $N_1=3$ and $N_2=2$, in line with standard flare studies \citep{Davenport_2016, Ilin_2021}. For the minimum duration criterion, we set $N_3=3$ for TESS and $N_3=5$ for CHEOPS, following \cite{Bruno_2024} and \cite{Poyatos_2025}, to reduce the potential false positives of residual magnetic activity in the CHEOPS light curves. This corresponds to minimum detectable durations of 1 minute for TESS and 15 seconds for CHEOPS. For each detected flare, \texttt{AltaiPony} recovers key flare parameters, including the flare amplitude, duration, start and stop times, and the ED. The ED is defined as the time integral of the flare flux relative to the quiescent level and represents the duration over which the star would need to emit at its quiescent luminosity to radiate the same energy as the flare. 

The detected flares were then examined for potential substructure using an iterative flare decomposition routine. We iteratively fitted one to five flare components using the \texttt{Llamaradas Estelares} flare model \citep{Mendoza_2022} and employing the \texttt{scipy.differential\_evolution} optimiser to determine the best-fit parameters. The Akaike information criterion (AIC) \citep{Akaike_1974} was computed for each iteration, based on the residuals calculated over a window of 3 minutes before and 15 minutes after the flare for TESS, and 30 seconds before to 150 seconds after for CHEOPS. Therefore, only flares sufficiently isolated from data gaps were retained. Following \cite{Davenport_2014}, we adopted the final model as the iteration that provides at least a 10\% improvement in AIC relative to the previous one.

\subsection{Energy and recovery probability correction}
\label{subsection:correction}
Flare detection is biased against low-energy events: large flares are almost always recovered, while low-energy flares are often missed. Furthermore, even after sigma-clipping, detrending can still be influenced by the presence of flares, inducing a flux decrease around them (see this effect in Figure 2 of \citet{Rajpurohit_2025}). This effect can affect the estimated flare energy, particularly for weaker events near the $3\sigma$ threshold used for sigma clipping. To quantify these biases, we performed a flare injection-recovery analysis following the methodology of \cite{Seli_2021} and \cite{Gao_2022}.

For each light curve, artificial flares were injected following the flare model, spanning relative amplitudes from $10^{-3}$ to $10^{0}$ and full width at half maximum (FWHM) durations from $10^{-2}$ to $10^1$ minutes. To avoid overlap, we injected one flare per day and ensured that at least one flare was injected in every light-curve segment. We performed 50 iterations per TESS light curve and 200 per CHEOPS light curve. Each injected light curve was detrended and processed with the same flare detection routines described in Sections \ref{subsection:detrending} and \ref{subsection:detection}. We define detection probability as the fraction of recovered to injected flares for a given energy. The energy recovery ratio is the average of the ratio between recovered and injected energy. These quantities were then used to assign detection probabilities and energy correction factors to each real flare. 

Alternative approaches, such as wavelet-based denoising, can be used to improve the recovery of low-energy flares. However, \cite{Poyatos_2025} showed that wavelets do not extend the observable flare energy range but enhance their recovery close to the detection threshold. In our analysis, this effect is already accounted for through the injection-recovery process, which quantifies detection probability as a function of flare energy and corrects for the resulting observing bias. Consequently, applying wavelets in addition to injection-recovery would not further alter the inferred slope of the FFD; therefore, we adopted the latter as our primary correction method.

To convert ED to flare energy, we followed the method of \cite{Shibayama_2013}, multiplying the quiescent stellar luminosity by the ED. As this energy is band-dependent, we applied bolometric correction factors to estimate the total energy emitted. Using the spectral response of each instrument and assuming a 10,000 K blackbody flare spectrum, we adopted $\epsilon_{CHEOPS}=0.34$ and $\epsilon_{TESS}=0.18$. Bolometric energy was computed as: 
\begin{equation}
    E_{bol} = L \times ED \times \epsilon^{-1}
\label{eq:ebol}
\end{equation}

\subsection{FFD construction and fit}
\label{subsection:ffd}

\begin{figure*}[b]
\centering
\includegraphics[width=0.9\textwidth]{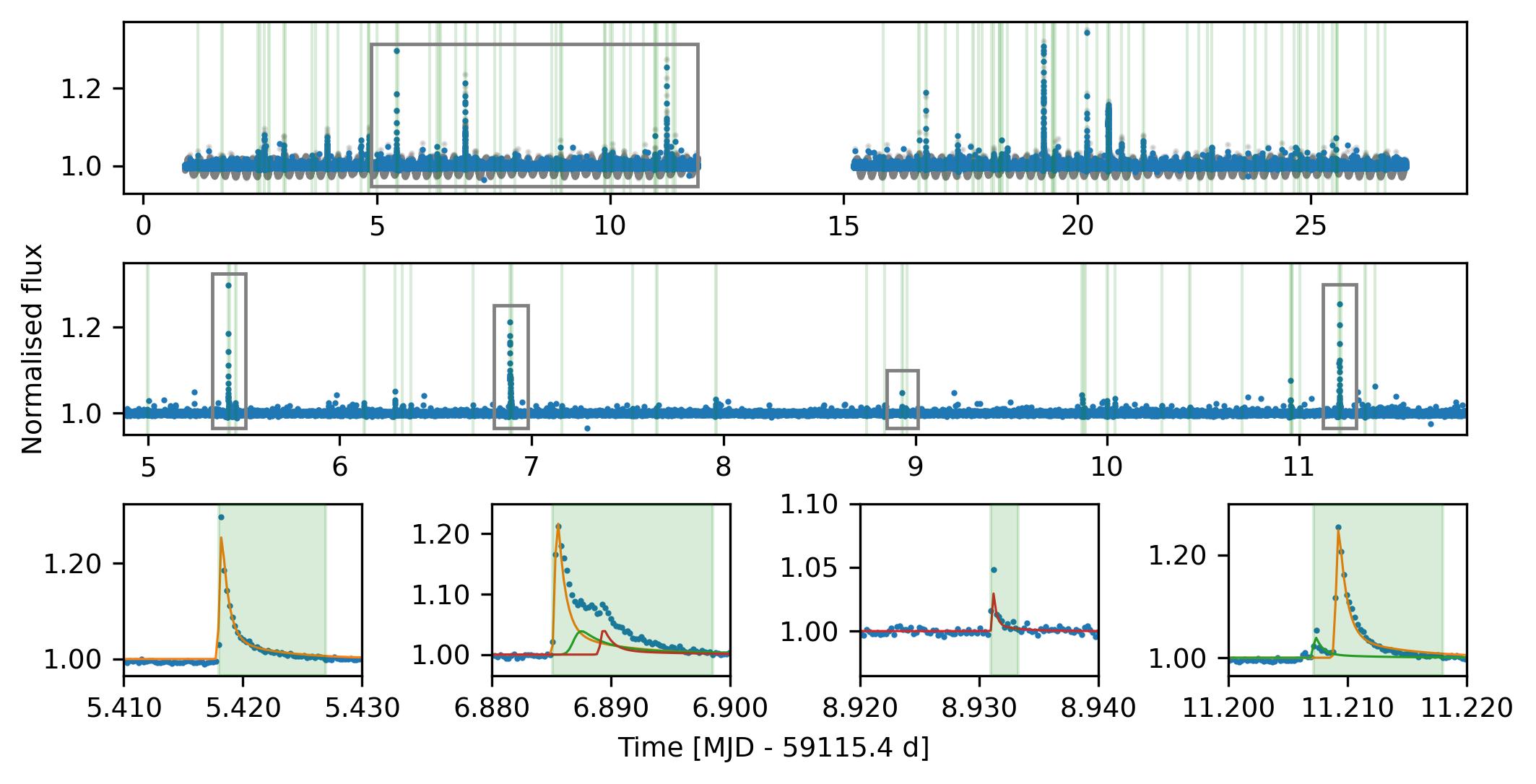}
\caption{Example TESS light curve of the M5V star GJ 65. The top panel shows the detrended light curve (blue points) from sector 30, with the raw light curve in the background (grey points). The middle panel presents a one-week zoom of a selected portion, while the bottom panel provides four close-ups of individual flares. Grey boxes mark the sections shown in the zooms. In the bottom panels, the fitted flare components are plotted as solid coloured lines.}
\label{fig:lightcurve_tess}
\end{figure*}

The FFDs describe the cumulative flare rate as a function of energy, representing the expected frequency of flares of a given energy or higher. In constructing FFDs, we followed the approach of \cite{Bruno_2024} by treating each fitted component of a complex flare obtained in Section \ref{subsection:detection} as an individual flare event. To account for star-to-star variations in detection thresholds, we applied the first-order correction implemented in \texttt{Altaipony} by setting the parameter \textit{multiple\_stars = True}, so each FFD segment was normalised by the number of contributing stars. We constructed FFDs separately for the TESS and CHEOPS flare samples, both before and after correcting for detection probability and flare energy recovery, as described in Section \ref{subsection:correction}. To enable fair comparison between the two missions, we scaled the number of flares detected with CHEOPS by the ratio of the TESS observing time to the CHEOPS observing time for each star. This accounts for the shorter CHEOPS time coverage and yields flare rates comparable to TESS. For the three targets without TESS observations, individual scaling factors could not be determined. In these cases, we instead applied a global scaling factor equal to the total TESS observing time divided by the total CHEOPS observing time, corresponding to $\sim$167.
We then constructed four sets of FFDs:
\begin{itemize}
    \item TESS flare frequency versus ED,
    \item CHEOPS flare frequency versus ED (scaled),
    \item TESS flare frequency versus bolometric energy,
    \item CHEOPS flare frequency versus bolometric energy (scaled).
\end{itemize}

Each FFD was fitted with multiple candidate distributions using the \texttt{curve\_fit} function from the \texttt{scipy.optimize} module. For the ED-based FFDs, we compared the fits using both a power-law and a lognormal distribution. For the $E_{\text{bol}}$-based FFDs, we additionally tested a truncated power-law model, which accounts for possible incompleteness or physical cutoffs at high energies. The truncation energies and the power-law indices on either side of the break were estimated using \texttt{curve\_fit} and the modified maximum likelihood estimator of \cite{Maschberger_2009}. Parameter uncertainties were estimated from the covariance matrix returned by \texttt{curve\_fit}: the square roots of its diagonal provide the standard errors of the fitted parameters, which we then scaled by the appropriate Student-t critical value to obtain two-sided 95\% confidence intervals. 

To determine the best-fitting model, we applied Vuong's likelihood ratio test \citep{Vuong_1989}. The candidate distributions (power law, lognormal, and truncated power law) are non-nested, meaning that none of the distributions can be expressed as a special case of another by simply constraining parameters. Because of this, the classical likelihood ratio test based on Wilks' theorem does not apply. For each distribution, we computed the Poisson log-likelihoods of the observed cumulative flare frequencies:
\begin{equation}
    l = \sum_{i=1}^{n} \Big[ y_i \log(\lambda_i) - \lambda_i - \log(y_i!) \Big]
\label{eq:loglikelihood}
\end{equation}
where $y_i$ is the observed number of flare events in bin $i$, and $\lambda_i$ is the expected flare count predicted by the candidate distribution. The normalised log-likelihood ratio statistic is then
\begin{equation}
    R = \frac{l_2 - l_1}{\sqrt{n\hat{\sigma}^2}}
\label{eq:ratio}
\end{equation}
where $\hat{\sigma}^2$ is the empirical variance of the per-point log-likelihood differences. Under the null hypothesis that both models fit equally well, $R$ is approximately standard normal. If $R$>0, the second distribution is preferred, while if $R$<0, the first distribution is preferred. The two-sided $p$ value was computed as 
\begin{equation}
    p = 2[1 - \Phi(|R|)]
\label{eq:p-value}
\end{equation}
where $\Phi$ is the cumulative distribution function of the standard normal. Lower $p$ values indicate stronger statistical support for the sign of $R$ and therefore for the preferred distribution.

Finally, we constructed a combined FFD by merging the CHEOPS and TESS samples. Before merging, each dataset was corrected for instrumental effects, bolometric energy, and detection biases, as is described in Section \ref{subsection:correction}. The CHEOPS flare counts were also scaled to match TESS observing times. This ensures that flare energies and frequencies from both missions are on a consistent scale, allowing us to treat the combined sample as if it were obtained from a single instrument, spanning a wider range of flare energies than either mission alone. As before, we fitted power-law and lognormal distributions to the ED-based FFD, and included a truncated power-law distribution for the $E_{\text{bol}}$-based FFD. The relative goodness of fit was assessed using the normalised log-likelihood ratio, $R$, and the corresponding $p$ value described previously, allowing us to determine the statistically preferred distribution in each case.

\section{Results}
\label{section:results}

\begin{figure*}
\centering
\includegraphics[width=0.9\textwidth]{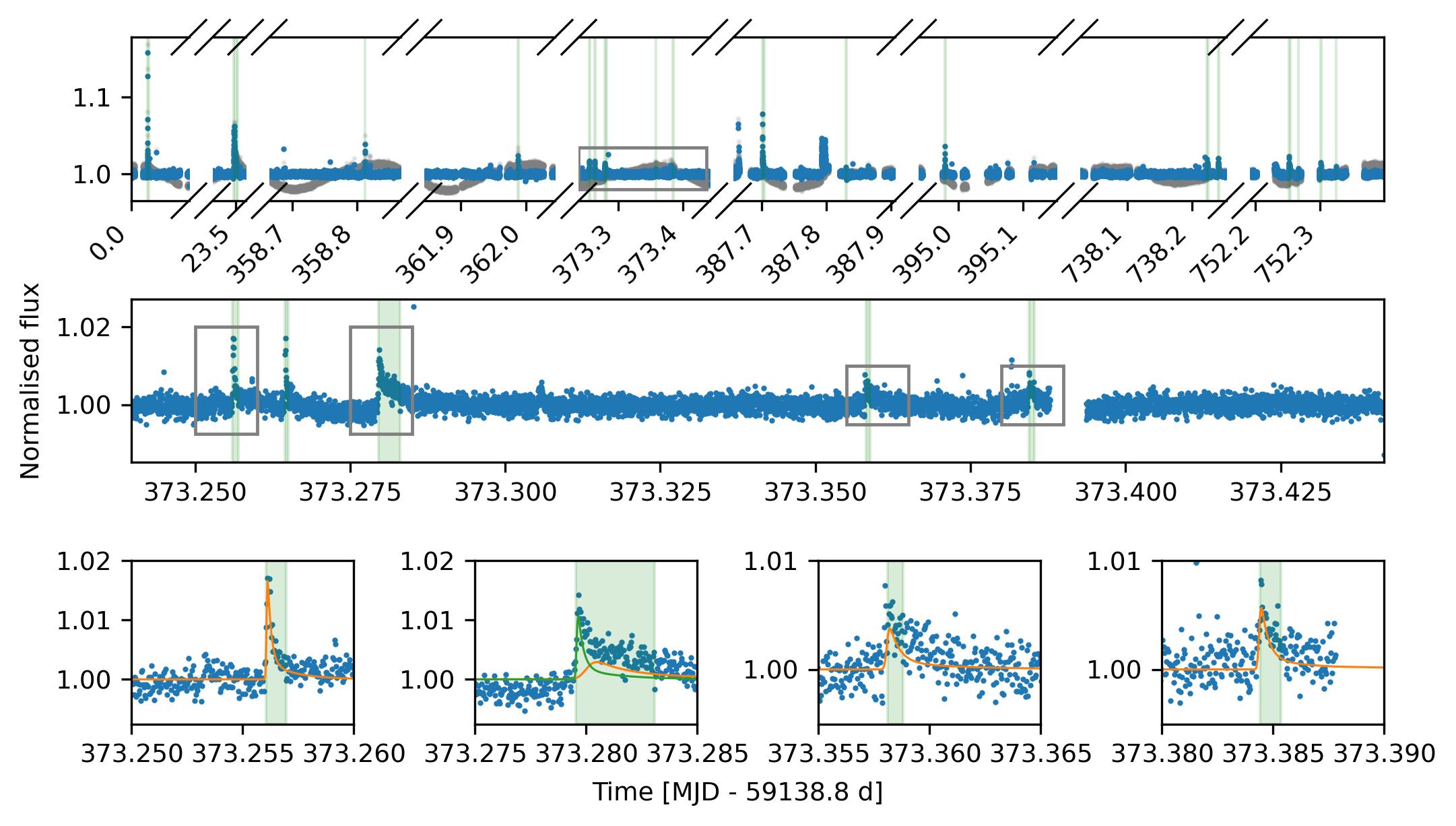}
\caption{Combined CHEOPS light curve of multiple visits of the M5V star GJ 65. The x axis in the top panel is broken to account for observational gaps. The top panel shows the detrended light curve (blue points) with the raw light curve in the background (grey points). The middle panel presents one full visit, while the bottom panel provides four close-ups of individual flares. Grey boxes mark the sections shown in the zooms. In the bottom panels, the fitted flare components are plotted as solid coloured lines.}
\label{fig:lightcurve_cheops}
\end{figure*}

We detected a total of 4150 flares in the TESS light curves of 78 stars and 179 flares in the CHEOPS light curves of 40 stars. Of these, 34 stars exhibited flares in both datasets. Following the flare decomposition methodology in Section \ref{subsection:detection}, we identified a large fraction of flares with complex structures. We found that 23\% of TESS flares and 44\% of CHEOPS flares are complex, consistent with previous studies \citep{Davenport_2014, Bruno_2024, Poyatos_2025}. After decomposition, the total number of flare events increased to 5,324 for TESS and 296 for CHEOPS. 

The higher fraction of complex flares detected by CHEOPS compared to TESS primarily results from the higher observational cadence. CHEOPS light curves typically have few-second cadences, providing more data points per flare, which increases the likelihood of resolving substructures within individual events. In contrast, TESS light curves have a 20- or 120-second cadence, limiting the temporal resolution and making some flares appear simpler. This cadence-dependent effect has been discussed in \cite{Howard_2022}, showing that binning 20-second flare light curves obtained by TESS on low-mass flare stars to a 120-second cadence produced degeneracies in complex flare structures (see their Figure 2). This is consistent with our results on M dwarfs, showing that CHEOPS identifies nearly twice as many complex flares as TESS.

\begin{figure}
\centering
\includegraphics[width=0.9\columnwidth]{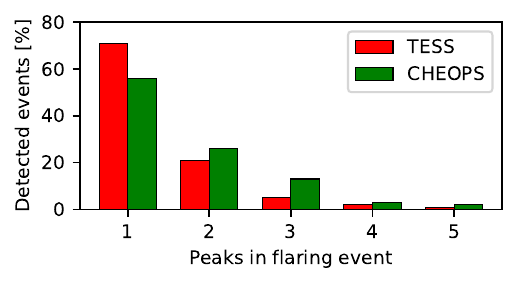}
\caption{Number of peaks recovered per flare event for TESS (red) and CHEOPS (green). Peaks correspond to individual flare components identified by the fitting procedure.}
\label{fig:peaks}
\end{figure}

Figure \ref{fig:lightcurve_tess} shows an example of a flare-rich TESS light curve of GJ 65, the target with the highest number of CHEOPS-detected flares, with simple and complex flare profiles highlighted. Figure \ref{fig:lightcurve_cheops} presents a CHEOPS light curve of the same star. In the bottom panels, the flare colours carry no physical meaning and simply indicate the order in which they were added during the fitting, which depends on the specific flare shape. We provide additional examples of complex flares in Figure \ref{fig:multiflare}. The distribution of the number of flare components per event across all light curves is shown in Figure \ref{fig:peaks}. The higher percentage of complex flares in CHEOPS was expected, given the instrument's superior cadence and sensitivity compared to TESS, consistent with findings from previous flare studies using CHEOPS \citep{Bruno_2024, Poyatos_2025}. However, some apparently simple flares may be intrinsically complex but with substructures that remain undetected due to the limited sensitivity or cadence of the instruments, as discussed by \citet{Howard_2022}. The reported fraction of components per flare should therefore be considered indicative, since it depends on the photometric precision and temporal resolution of each mission.

\begin{figure}
\centering
\includegraphics[width=0.9\columnwidth]{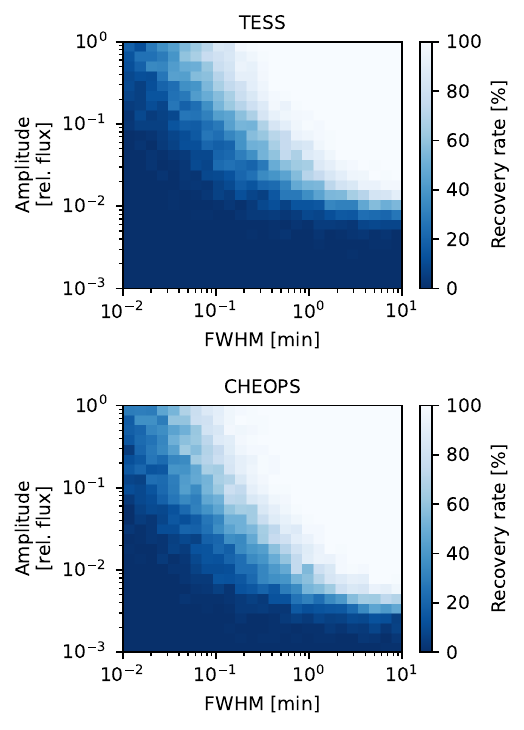}
\caption{Flare recovery rate from synthetic flare injection-recovery tests for TESS (top) and CHEOPS (bottom).}
\label{fig:recovery}
\end{figure}

This sensitivity advantage is illustrated in Figure \ref{fig:recovery}, which presents the results of the flare injection-recovery process described in Section \ref{subsection:correction}. For flares with identical injected amplitudes and durations (FWHM), CHEOPS consistently achieves recovery rates higher than TESS. This is attributed to its higher photometric precision and faster cadence. 

\begin{figure}[t]
\centering
\includegraphics[width=0.9\columnwidth]{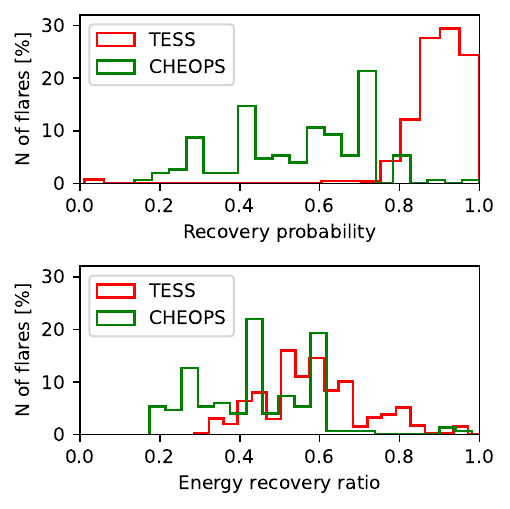}
\caption{Histograms of the obtained recovery probabilities (top) and energy recovery ratios (bottom) for TESS (red) and CHEOPS (green).}
\label{fig:histograms}
\end{figure}

Figure \ref{fig:histograms} displays the resulting detection probabilities and energy recovery ratios for the detected flares. While CHEOPS is more sensitive overall, it also exhibits a broader range of recovery probabilities, indicating that it detects more weak flares but also misses a greater fraction of them compared to high-energy ones. In contrast, TESS fails to recover the weakest flares detected by CHEOPS altogether, showing mostly a recovery probability above 0.8, which indicates a strong observational bias at lower flare energies. This highlights the unique contribution of CHEOPS in probing the low-energy end of the flare distribution. The energy recovery ratio is similar across both missions, suggesting that the detrending process biases comparably flare energy estimates. Notably, we do not observe energy overestimation (ratios > 1) as reported in \cite{Gao_2022}.

\begin{figure}
\centering
\includegraphics[width=0.9\columnwidth]{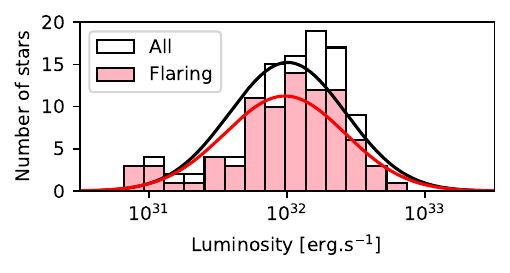}
\caption{Histogram of quiescent luminosities for all stars (white) and flaring stars (red), with lognormal fits shown as black and red lines, respectively.}
\label{fig:luminosities}
\end{figure}

We estimated the quiescent bolometric luminosity of each target using its radius and effective temperature. Figure \ref{fig:luminosities} presents the luminosity distributions for all stars and for the subset with detected flares. Both distributions are roughly lognormal, with $\mu=32.00, \sigma=0.42$ for the full sample and $\mu=31.98, \sigma=0.44$ for the flaring stars. However, we note substantial statistical uncertainty because of the relatively small sample size of 110 stars. 

\begin{figure*}
\centering
\includegraphics[width=0.85\textwidth]{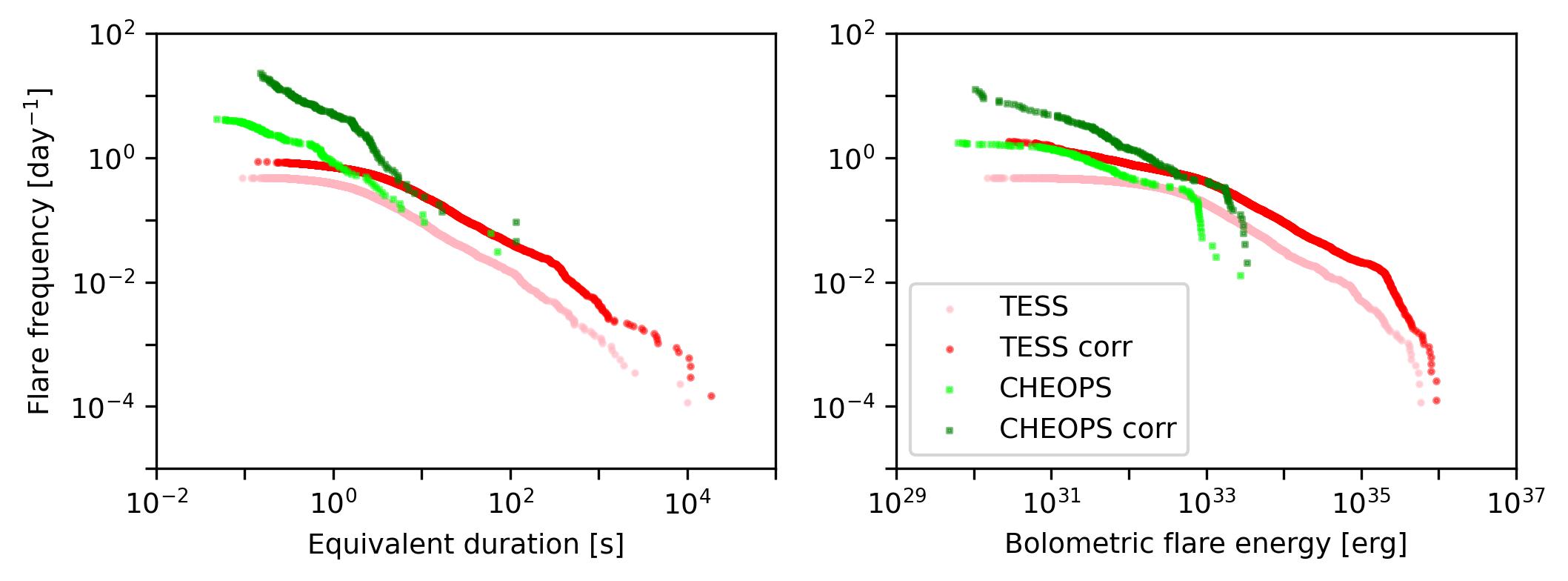}
\caption{Flare frequency distributions for the TESS sample (light and dark red) and the CHEOPS sample (light and dark green), shown before and after corrections, with ED-based FFDs in the left panel and $E_{\text{bol}}$-based FFDs in the right panel.}
\label{fig:ffds}
\end{figure*}

We constructed FFDs in terms of both ED and bolometric energy, as described in Section \ref{subsection:ffd}, and the resulting distributions are shown in Figure \ref{fig:ffds}. The CHEOPS-based FFDs extend to lower energies and ED values than those from TESS, consistent with CHEOPS's ability to detect weaker flares. The FFDs from the two missions overlap, with the upper end of the CHEOPS distribution aligning with the lower end of the TESS distribution. This overlap suggests a coherent extension of the flare distribution across both datasets. 

\begin{figure*}
\centering
\includegraphics[width=0.85\textwidth]{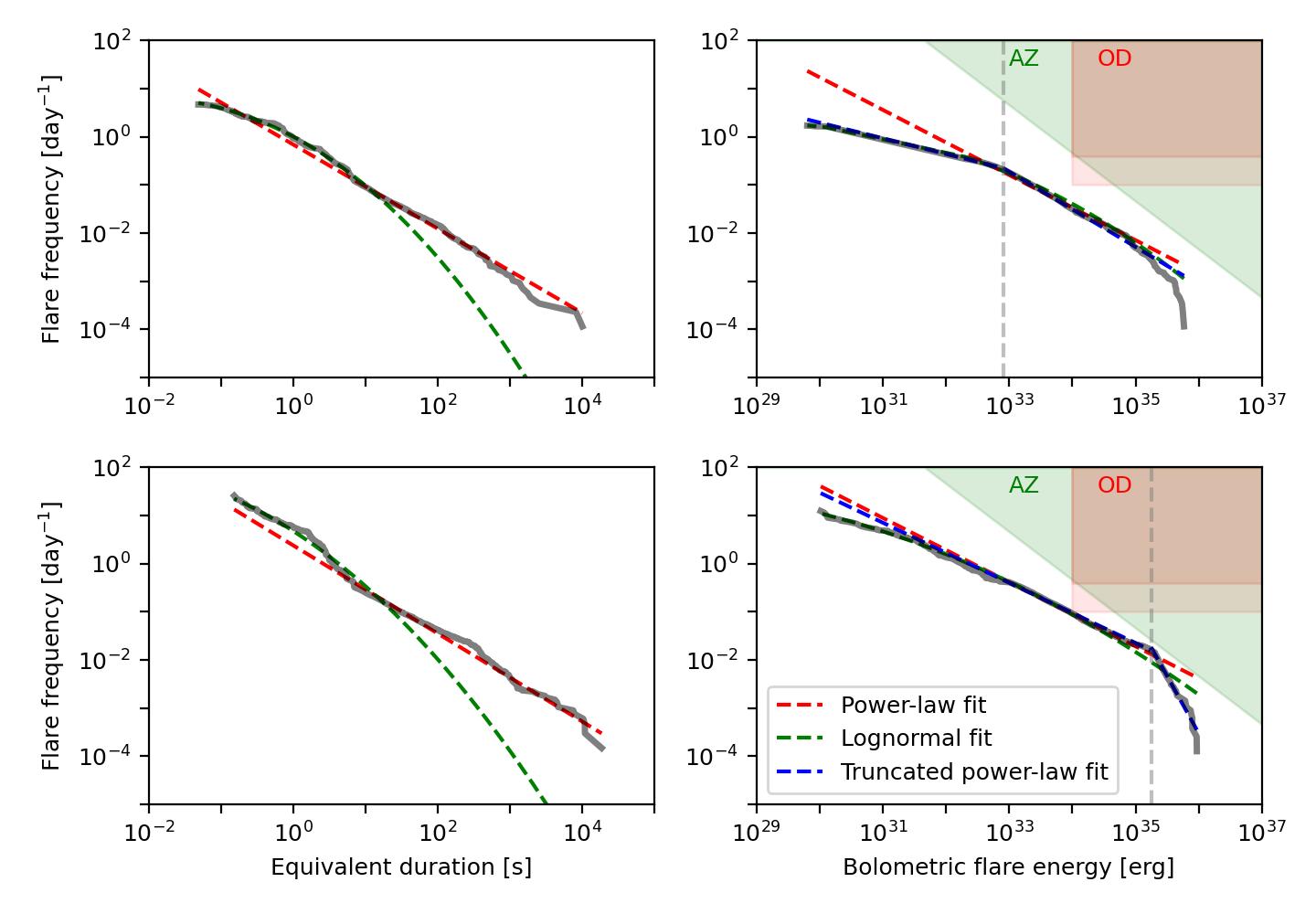}
\caption{Combined FFDs (grey) for the observed (top) and corrected (bottom) samples. ED-based FFDs are shown in the first and second panels, and $E_{\text{bol}}$-based FFDs in the third and fourth panels. Power-law and lognormal fits are shown with dashed red and green lines, respectively, while truncated power-law fits for the $E_{\text{bol}}$-based FFDs are shown with dashed blue lines. The dashed vertical grey lines mark the truncation thresholds $t_1$ and $t_2$. The shaded green areas correspond to the abiogenesis zone (AZ) \citep{Rimmer_2018}. The shaded light and dark red areas correspond to the permissive and conservative thresholds of the ozone depletion zone (OD) \citep{Tilley_2019}, respectively.}
\label{fig:fits}
\end{figure*}

We then fitted all FFDs, both per-mission and combined, following the methodology described in Section \ref{subsection:ffd}. The best-fit parameters, the normalised log-likelihood ratios $R$ and the associated values $p$ are reported in Table \ref{table:fits}, while Figure \ref{fig:fits} shows the combined FFDs with the fitted distributions. After decomposing complex flares and correcting for detection biases, the FFDs become steeper, corresponding to an increase in the fitted $\alpha$ index and reflecting a higher inferred frequency of weaker flares. This effect is strongest for CHEOPS, where the corrected FFD rises sharply at low energies because of the strong observational bias against short, low-amplitude events. For ED-based FFDs, the $\alpha$ index increases from $\alpha=1.67\pm 0.02$ to $\alpha=1.95\pm 0.03$, and for $E_{\text{bol}}$-based FFDs from $\alpha=1.45\pm 0.04$ to $\alpha=1.71\pm 0.02$. For TESS, the ED-based FFD shows no significant change, with $\alpha$ varying from $1.66\pm 0.01$ to $1.65\pm 0.01$, while the $E_{\text{bol}}$-based FFDs exhibits a modest increase from $1.48\pm 0.01$ to $1.51\pm 0.01$. For the combined FFDs, the ED-based distributions steepen from $1.85\pm 0.01$ to $1.90\pm 0.01$, and the $E_{\text{bol}}$-based ones from $1.63\pm 0.01$ to $1.70\pm 0.01$. After fitting a truncated power law on the $E_{bol}$-based FFDs, we find that, for the observed sample, the $\alpha$ index changes from $1.30 \pm 0.02$ to $1.78 \pm 0.03$, with a truncation threshold at $t_1 = 4.8 \times 10^{32}$ erg. For the corrected sample, the $\alpha$ index increases from $1.66 \pm 0.02$ to $3.39 \pm 0.07$, with a corresponding truncation threshold of $t_2 = 1.8 \times 10^{35}$ erg.

For per-mission FFDs, the power-law distribution generally provides a better fit for the ED-based distributions, while the lognormal distributions give a better description of the $E_{bol}$ distributions, with the exception of the corrected CHEOPS sample. For the combined ED-based distributions, a power law also provides the best fit, although the associated $p$ value exceeds 0.05 for the observed sample. In contrast, the combined distributions based on bolometric energy are better described by a lognormal distribution, though in this case the associated $p$ value also exceeds 0.05 for the corrected sample. However, a truncated power law provides an even better fit, consistent with the results of \cite{Davenport_2014}, although the associated value of $p$ remains above 0.05. The corresponding truncation energy is estimated to be $1.8 \times 10^{35}$ erg, which may reflect either the limited TESS observing baseline, insufficient to accurately sample the rarest, highest-energy events, or a genuine change in flare physics at the highest energies.

\begin{table*}[htp]
\caption{Fit parameters for per-mission and combined FFDs, both ED- and $E_{\text{bol}}$-based. }
\begin{adjustbox}{width=\textwidth}
\begin{tabular}{lccccccccc}
\hline
     FFD                       & Power law             & Lognormal            & Truncated            & R$_{PL}$               & p$_{PL}$          & R$_{LT}$               & p$_{LT}$    & $\mathrm{R}_{PT}$               & $\mathrm{p}_{PT}$       \\ \hline
TESS (ED)                   &      $\alpha=1.66$                &       $\mu=0.43, \ \sigma=0.61$               &     --                 &       -6.24               &      < 0.001                &       --               &         --             \\
TESS corr. (ED)         &     $\alpha=1.65$                 &       $\mu=0.63, \ \sigma=0.69$               &     --                 &      -7.81                &     < 0.001                &        --              &          --            \\
CHEOPS (ED)                 &    $\alpha=1.67$                  &       $\mu=-1.24, \ \sigma=1.00$               &    --                  &         -0.60             &      0.055             &       --               &         --             \\
CHEOPS corr. (ED)       &    $\alpha=1.95$                  &       $\mu=-14.88, \ \sigma=2.85$             &   --                   &       -1.87              &      0.061               &       --               &         --             \\
TESS (bol. E)               &      $\alpha=1.48$                &      $\mu=32.80, \ \sigma=0.74$                &             --         &         1.65           &     0.098                   &        --              &         --             \\
TESS corr. (bol. E)     &     $\alpha=1.51$                 &      $\mu=31.50, \ \sigma=1.56$                &             --         &     1.11                &       0.267                &         --             &             --         \\
CHEOPS (bol. E)             &      $\alpha=1.45$                &   $\mu=31.53, \ \sigma=0.86$                   &              --        &       2.40               &     0.016                   &    --                  &        --              \\
CHEOPS corr. (bol. E)   &       $\alpha=1.71$               &      $\mu=27.80, \ \sigma=1.79$                  &            --          &    -1.36                  &     0.173                  &      --                &       --               \\
Combined (ED)               & $\alpha=1.85$ & $\mu=-1.14, \ \sigma=0.95$ & -- & -2.72 & 0.066 & -- & -- \\
Combined corr. (ED)     & $\alpha=1.90$ & $\mu=-2.25, \ \sigma=1.10$ & -- & -2.02 & 0.043 & -- & -- \\
Combined (bol. E)           & $\alpha=1.63$ & $\mu=31.37, \ \sigma=1.21$ & $\alpha_{<t_1}=1.30, \ \alpha_{>t_1}=1.78$ & 5.09 & 0.039 & 4.96 & 0.072 & 9.15 & 0.022 \\
Combined corr. (bol. E) & $\alpha=1.70$ & $\mu=27.95, \ \sigma=1.75$ & $\alpha_{<t_2}=1.66, \ \alpha_{>t_2}=3.39$ & 0.44 & 0.066 & 2.30 & 0.060 & 4.43 & 0.034 \\ \hline
\label{table:fits}
\end{tabular}
\end{adjustbox}
\vspace{-0.5em}
\tablefoot{Columns show the power-law $\alpha$, lognormal $\mu$ and $\sigma$, $\alpha$ below and above $10^{33}$ erg if a truncated power law is applicable, and normalised likelihood ratios with $p$ values comparing the power-law and lognormal fits, the lognormal and truncated power-law fits, and the power-law and truncated power-law fits.}
\end{table*} 

\begin{figure*}[h]
\centering
\includegraphics{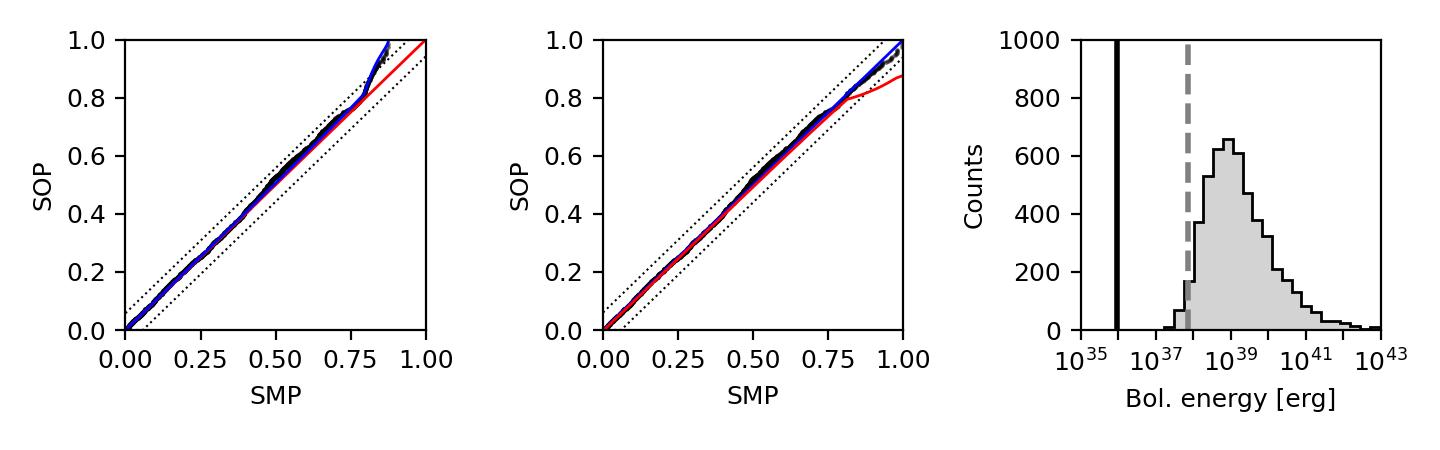}
\caption{Left: Stabilised percentile-percentile plot for the null hypothesis of the best-fit infinite power law (red diagonal), showing stabilised model percentiles (SMPs) versus stabilised observed percentiles (SOPs). The corrected sample is shown as black data points. Percentiles of the best-fit truncated power law are indicated by blue line, and the 5\% significance-level stabilised Kolmogorov-Smirnov acceptance region is shown by the two lines parallel to the diagonal. Middle: Stabilised percentile-percentile plot for the null hypothesis of the best-fit truncated power law. Right: Exceedance test results, comparing the observed maximum energy (vertical black line) with the distribution of simulated maxima (grey histogram). The 2.5th percentile of the simulated maxima (dashed grey line) is derived from 5000 trials drawn from the best-fit infinite power law.}
\label{fig:sks}
\end{figure*}

The $E_{\text{bol}}$-based FFDs in Figure \ref{fig:fits} also show regions relevant to planetary habitability, where ozone depletion or abiogenesis may occur. \cite{Tilley_2019} estimates that the ozone layer of unmagnetised planets can be completely destroyed if the superflare rate with energies above $10^{34}$ erg reaches 0.4 per day, with a more permissive limit at 0.1 per day. In contrast, \cite{Rimmer_2018} originally defined the lower limit of the FFD required for abiogenesis in terms of U-band flare energy, following a power law with $\alpha = 2$. Here, we adopted Equation 10 of \cite{Gunther_2020} to convert this U-band threshold to bolometric flare energy, which depends on the stellar radius and effective temperature. We estimated that $\sim$7.8\% of the energy of a 10,000K flare blackbody falls within the U-band, and determined the mean radius and effective temperature of the 84 active stars in our sample as $\sim0.456 R_\odot$ and $\sim3,525$K, respectively. For our observed sample, the slope of the high-energy segment of the truncated power-law distribution, which describes superflare frequencies, is approximately parallel to the abiogenesis zone boundary, suggesting that the two would not intersect even if the high-energy bias were corrected. However, in the corrected sample, this parallelism disappears, indicating that correcting the high-energy bias with longer baselines would allow a more reliable assessment of the impact of superflares on planetary habitability to be undertaken.

To better assess deviations in the high-energy bins, which contain only a small number of events as a result of extreme-value statistics, we applied the variance-stabilising transformations described by \cite{Maschberger_2009} (Equations 18 and 19) to emphasise the behaviour of the distribution tail. We then constructed the stabilised percentile-percentile (SPP) plots shown in Figure \ref{fig:sks} by comparing the corrected flare frequencies with those expected from the best-fit infinite and truncated power laws. The two SPP plots also include the acceptance region of the stabilised Kolmogorov-Smirnov (sKS) statistic at the 5\% significance level. We find that, at the highest percentiles, the corrected distribution deviates from the infinite power law and exits the sKS acceptance region, allowing us to reject the null hypothesis of an infinite power law. In contrast, it remains within the sKS acceptance region for the truncated power law, so this null hypothesis cannot be rejected.

To determine whether the observed truncation arises from an intrinsic physical process or from sampling effects, we followed the exceedance test recommended by \cite{Maschberger_2009}. Specifically, we drew the same number of data points (5620) from the best-fit infinite power law, recorded the maximum energy of each trial, and repeated this procedure 5000 times. We then compared the maximum energy observed in our data set with the 2.5th percentile of the resulting distribution of simulated maxima, as is shown in the right panel of Figure \ref{fig:sks}. Because the observed maximum lies below the 2.5th percentile, and assuming that the truncated power law is the simplest explanation, we suggest that the observed high-energy deviation from an infinite power law is more likely due to limited sampling than to an intrinsic physical truncation.

To assess the impact of sample heterogeneity on our results, we repeated the analysis for an active subsample composed of the 12 stars with more than 100 detected flares; namely, AD Leo, AU Mic, EQ Peg, EV Lac, G 99-49, GJ 358, GJ 494, GJ 65, Gl 799B, Gl 841 A, Proxima Cen, and Ross 733. The corresponding corrected FFDs, SPP plots, and exceedance test results are presented in Appendix C. As with the full sample, the infinite power-law hypothesis is rejected at high percentiles, while the truncated power-law hypothesis is not, and the exceedance test suggests that the observed deviation can be attributed to limited sampling.

\section{Discussion}
\label{section:discussion}
The FFDs are often assumed to follow a power law. However, when the energy range under study is sufficiently broad, deviations from a pure power law are frequently observed \citep{Verbeeck_2019, Howard_2019}. These deviations may arise from observational biases or reflect intrinsic properties of flare generation that diverge from ideal scaling laws. 

We found that FFDs constructed from ED (the quantity directly measured from photometry) are well described by a power law. This suggests that, observationally, flare occurrence scales predictably with ED. In contrast, when ED is converted into bolometric energy, the distributions deviate from a simple power law because of the influence of stellar luminosity. Since stellar luminosities in our sample follow an approximately lognormal distribution (see Figure \ref{fig:luminosities}), the convolution of a power-law ED distribution with this lognormal luminosity distribution yields an energy-based FFD that no longer follows a pure power law. This result is consistent with the findings of \cite{Bruno_2024} and \cite{Poyatos_2025}, who similarly concluded that FFDs cannot be adequately described by either a power law or lognormal distribution alone. However, the corrected FFD obtained by \cite{Seli_2021} still closely follows a power law. This likely results from their relatively homogeneous sample of TRAPPIST-1 analogue stars, so the quiescent luminosity distribution introduces less distortion into the resulting energy-based FFDs (see their Figure 21). For our sample, the best representation is a truncated power law with a break around $10^{35}$ erg. 

By combining CHEOPS and TESS and correcting for flare recovery completeness and energy underestimation, we extend the sensitivity of the FFDs between $10^{30}$ and $10^{33}$ erg. This mitigates the dominant observational bias causing the apparent flattening of the FFD at low energies, thereby reinforcing the power-law slope. Notably, this addresses one of the main motivations behind previous attempts to fit lognormal models, namely the apparent underrepresentation of weak flares. Our result demonstrates that, with sufficiently sensitive missions and robust flare decomposition, the low-energy regime can be reconciled with a power-law distribution.

\begin{figure}
\centering
\includegraphics[width=0.9\columnwidth]{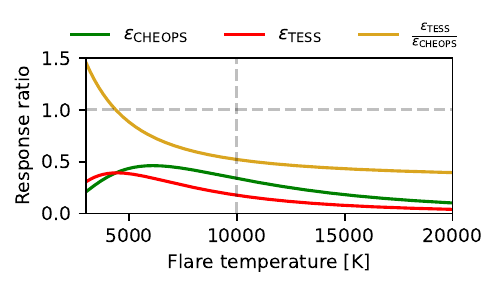}
\caption{Evolution of the $\epsilon_{CHEOPS}$ (green) and $\epsilon_{TESS}$ (red) bolometric conversion factors as a function of the assumed flare blackbody temperature. The ratio $\epsilon_{TESS} / \epsilon_{CHEOPS}$ is shown in yellow.}
\label{fig:blackbody_ratio}
\end{figure}

A key step in comparing TESS and CHEOPS FFDs involves accounting for each mission's response to a flare spectrum. A blackbody temperature of 10,000 K is commonly assumed for M dwarf flares \citep{Kowalski_2024}, but observed flare temperatures span 5,000 - 30,000 K \citep{Howard_2020, Maas_2022, Bicz_2025}. Selecting a flare blackbody temperature based only on optical photometry can provide large gaps in the calculated energy \citep{Jackman_2023, Berger_2024}. Because TESS and CHEOPS passbands differ (see Figure \ref{fig:blackbody}), the assumed flare temperature directly affects the bolometric correction factors $\epsilon_{TESS}$ and $\epsilon_{CHEOPS}$ (Section \ref{subsection:correction}), and thus the alignment of the two FFDs on the energy axis. Importantly, the assumed temperature affects only the bolometric energy, not the flare frequencies, so the impact is restricted to the horizontal scaling of the distributions. As is shown in Figure \ref{fig:blackbody_ratio}, the ratio $\epsilon_{TESS}/\epsilon_{CHEOPS}$ stabilises near 0.5 for temperatures between 10,000 K and 20,000 K, implying that our results are robust to temperature uncertainties over this range. In contrast, adopting cooler temperatures would reduce the overlap between the TESS and CHEOPS distributions and could introduce artificial offsets. 

At the high-energy end of the FFDs, additional uncertainty persists due to extreme-value statistics. We show that the apparent suppression of flares above $10^{35}$ erg is primarily driven by the limited sampling, as TESS, with observational baselines of at least 27 days, is unlikely to accurately capture events occurring on timescales of months to years. For comparison, \cite{Vasilyev_2024} used Kepler photometry to show that Sun-like stars produce superflares with energies exceeding $10^{34}$ erg at a rate of about once per century, consistent with an extrapolation of the solar flare distribution to higher energies. Correcting for sampling bias at the highest energies is therefore essential for determining whether the mechanisms responsible for generating these energetic superflares differ from those that trigger lower-energy flares. Alternative models, such as the tapered power-law distribution proposed by \cite{Sakurai_2022}, suggest that the flare production may decline sharply beyond a characteristic energy scale, reflecting a maximum attainable flare energy that depends on the size of the stellar convection zone.

A physical deviation from power-law behaviour at high energies would have important consequences for exoplanet habitability. Large flares ($\gtrsim10^{34}$ erg) have the strongest impact on planetary atmospheres, driving processes such as the depletion of atmospheric compounds (e.g., ozone and methane) and the initiation of prebiotic chemistry through enhanced UV radiation \citep{Segura_2010, Rimmer_2018, Tilley_2019}. If the true flare distribution deviates from a pure power law, the occurrence rate of the most energetic events would be lower than previously estimated. As a consequence, the cumulative flare fluence experienced by planets orbiting active M dwarfs may be reduced, potentially lowering the expected frequency of extreme atmospheric disruption. This would affect models of exoatmospheres in M-dwarf systems, particularly within habitable zones where flare-driven processes play a critical role \citep{Konings_2022}.

Furthermore, accurate modelling of FFDs is crucial not only for assessing planetary habitability but also for designing and evaluating future flare-monitoring missions. Flare distributions are commonly used as input for synthetic light-curve generation and recovery simulations. If the underlying FFD model is biased due to incomplete flare recovery or oversimplified parameter assumptions, predictions for flare detectability, cadence requirements, and mission sensitivity may be skewed. Our combined and corrected approach, which merges CHEOPS and TESS data and applies injection-recovery and decomposition corrections, provides a framework for generating more realistic flare populations for mission planning. Implementing corrected FFDs will yield more reliable mission simulations and better-informed observational strategies.

Our statistical assessment of the preferred FFD model relied on Vuong's likelihood ratio test, which is appropriate for comparing non-nested distributions. However, this method provides only a relative measure of support, and the associated $p$ values are sensitive to the sample size (see the high $p$ values obtained for the CHEOPS FFDs in Table \ref{table:fits}). Alternative approaches could complement this analysis. For example, confidence intervals on fitted parameters, bootstrap resampling of the flare sample, or Bayesian model comparison using Bayes factors or posterior predictive checks could help quantify model stability and provide insight into the robustness of the inferred distributions. While implementing these methods is beyond the scope of the present work, future studies with larger flare samples would benefit from applying multiple statistical diagnostics in parallel. 

Finally, synergy between different missions remains a critical component in advancing FFD studies. Beyond combining TESS and CHEOPS, Kepler's long continuous light curves have been used to characterise flare occurrence \citep{Gao_2022}. Although only a small number of CHEOPS targets fall within the Kepler field, using the partial overlap and extending such analyses with Kepler observations would enable more consistent cross-mission comparisons and bridge gaps between baselines and sensitivities. In addition, ground-based surveys such as NGTS \citep{Wheatley_2018} and Evryscope \citep{Law_2015} can complement space missions by monitoring large samples of stars over long timescales. While their photometric precision is generally lower than that of space telescopes such as CHEOPS and Kepler, NGTS can achieve a precision comparable to TESS when operating in multi-telescope mode for selected targets. Their wide-field, high-duty-cycle observations also provide valuable statistics on the most energetic flares \citep{Howard_2019, Jackman_2021}. Combining FFDs from various missions would also provide a preliminary step ahead of the upcoming PLATO mission. PLATO will further improve our understanding, thanks to its wide sky coverage, multi-year observational baselines, high cadence, and precise photometry \citep{Rauer_2025}. We expect PLATO to increase the number of detected flares across the full energy range, help separate observational limitations from the intrinsic flare generation, and better inform studies of planetary atmospheres.

\section{Conclusion}
\label{section:conclusion}
We investigated the FFDs of M dwarf stars using complementary datasets from TESS and CHEOPS. By combining the long TESS baselines with the high photometric precision and cadence of CHEOPS, we extended the dynamic range over which stellar flares can be detected and characterised. This allowed us to probe both the low- and high-energy regimes of the FFD with greater accuracy than single-mission studies. 

We developed a comprehensive flare-analysis pipeline including detection, decomposition, injection-recovery tests, and corrections of flare frequency and bolometric energy. This enabled us to construct FFDs based on ED and bolometric energy while accounting for observational biases and instrument-specific sensitivities. Our results show that ED-based FFDs are well described by a power law, consistent with expectations from flare scaling laws. However, when converted into bolometric energy, the FFDs deviate from a simple power law due to the convolution with the lognormal luminosity distribution of the sample. %We concluded that this deviation, which appears above $10^{35}$ erg, is primarily caused by the limited sampling of such energetic flares.

We found that a truncated power law with a slope change around $10^{35}$ erg provides the best fit to $E_{\text{bol}}$-based FFDs, though statistical significance was achieved only at $p$ value < 0.1. This reflects two key features of the distributions. First, the apparent flattening at low energies, which has previously motivated the use of lognormal models, is shown here to result from observational biases of low-energy flares. After correcting for these effects and combining TESS and CHEOPS, we show that the resulting FFDs follow a power law up to $\sim 10^{35}$ erg. Beyond this threshold, the distribution deviates from a pure power law. After conducting stabilised Kolmogorov-Smirnov and exceedance tests, we attribute this deviation to limited sampling of the most energetic events.

The implications of these findings are twofold. First, for exoplanet habitability, a truncated power law suggests that superflares, though impactful, occur less frequently than previously assumed under pure power law extrapolations. This could slightly relax concerns about atmospheric erosion and biological sterilisation on planets orbiting active M dwarfs. Second, for future missions, our corrected and extended FFDs provide more realistic inputs for flare recovery simulations, improving mission planning and sensitivity estimates. 

Ultimately, this study demonstrates the value of combining multiple missions to overcome single-dataset limitations and improve our understanding of stellar flare statistics. As future missions such as PLATO become operational, offering longer baselines and improved precision, remaining uncertainties (especially at the highest energies) can be further constrained, refining models of stellar activity and its broader astrophysical implications. 

\section{Data availability}

Table \ref{annex:table} is available in electronic form at the CDS via anonymous ftp to \href{cdsarc.u-strasbg.fr}{cdsarc.u-strasbg.fr}(130.79.128.5) or via \href{http://cdsweb.u-strasbg.fr/cgi-bin/qcat?J/A+A/}{http://cdsweb.u-strasbg.fr/cgi-bin/qcat?J/A+A/}.

\begin{acknowledgements}
We thank the anonymous referee for their comments which helped improved the clarity of the paper. This work was partially supported by the Spanish programme MICIN/AEI/10.13039/501100011033, by the “European Union Next Generation EU/PRTR” through grant PCI2022-135049-2, by the “ERDF A way of making Europe” by the European Union through grants PID2021-125627OB-C31 and PID2022-136828NB-C41, by the programme Unidad de Excelencia María de Maeztu CEX2024001451-M to the ICCUB, and grants 2021SGR00679 and 2021SGR01108 by the Generalitat de Catalunya SGR-Cat 2021 programme. 
CHEOPS is an ESA mission in partnership with Switzerland, with important contributions to the payload and the ground segment from Austria, Belgium, France, Germany, Hungary, Italy, Portugal, Spain, Sweden, and the United Kingdom. CHEOPS public data analysed in this article are available in the CHEOPS mission archive.
This paper includes data collected by the TESS mission. Funding for the TESS mission is provided by NASA's Science Mission Directorate.
Additional resources include the SIMBAD database and VizieR catalog access tool from CDS, Strasbourg Astronomical Observatory, France \citep{Wenger_2000, Ochsenbein_2000}.
\end{acknowledgements}

\bibliographystyle{aa}
\bibliography{aanda}

\appendix

\onecolumn    
\begin{landscape}
\section{Target table}

\setlength{\tabcolsep}{1pt}  % default is 6pt
\begin{longtable}{l 
                  c 
                  >{\centering\arraybackslash}p{1.6cm} 
                  c 
                  >{\centering\arraybackslash}p{1.5cm} 
                  >{\centering\arraybackslash}p{1.5cm} 
                  >{\centering\arraybackslash}p{1.8cm} 
                  >{\centering\arraybackslash}p{2.2cm} 
                  >{\centering\arraybackslash}p{2.5cm} 
                  >{\centering\arraybackslash}p{2.8cm} 
                  >{\centering\arraybackslash}p{1.6cm} 
                  >{\centering\arraybackslash}p{1.8cm}}
\caption{Full table of targets, stellar parameters, and flare counts.} \\
\hline
Target & 
TIC ID & 
Spectral\newline type & 
Gmag & 
Teff\newline[K] & 
Dist\newline[pc] & 
Radius\newline[\(\textup{R}_\odot\)] & 
V sin i\newline[km s$^{-1}$] & 
TESS obs.\newline time [d] & 
CHEOPS obs.\newline time [d] & 
TESS\newline flares & 
CHEOPS\newline flares \\
\hline
\endfirsthead

\multicolumn{12}{c}{{\bfseries Table continued from previous page}} \\
\hline
Target & 
TIC ID & 
Spectral\newline type & 
Gmag & 
Teff\newline[K] & 
Dist\newline[pc] & 
Radius\newline[\(\textup{R}_\odot\)] & 
V sin i\newline[km s$^{-1}$] & 
TESS obs.\newline time [d] & 
CHEOPS obs.\newline time [d] & 
TESS\newline flares & 
CHEOPS\newline flares \\
\hline
\endhead

\hline \multicolumn{12}{r}{{Continued on next page}} \\
\endfoot

\hline
\endlastfoot

% -------------------
% Table rows
% -------------------
2MASS J03413724+5513068 & 450182870  & M2V & 10.55 & 4050.00 & 35.85 & 0.663 & 4.5 & 59.55 & 0.03 & 66 & \\
2MASS J06144242+4727346 & 307338124  & M0V & 10.81 & 3739.48 & 37.35 & 0.619 & 7.3 & 18.13 & 0.15 & & \\
2MASS J06192947+1357031 & 438007268  & M0V           & 10.01 & 3739.48      & 25.08         & 0.61              & 1.0                  & 87.50                          & 0.15                            &             & 2             \\
2MASS J09304457+0019214 & 383218397  & M3V           & 10.49 & 3275.05      & 9.9           & 0.323             & 1.6                  & 78.71                         & 0.13                            & 11          & 5             \\
2MASS J11421839+2301365 & 119584412  & M0V           & 10.83 & 3739.48      & 30.71         & 0.546             & 1.0                  & 41.01                         & 0.07                            &             &               \\
2MASS J11474440+0048164 & 325275315  & M3V           & 9.59  & 3122.25      & 3.37          & 0.21              & 3.7                  & 47.26                         & 0.18                            &             &               \\
2MASS J13314666+2916368 & 368129164  & M4V           & 10.61 & 3122.25      & 18.29         & 0.539             & 55.8                 & 55.82                         & 0.2                             & 90          & 6             \\
2MASS J20103444+0632140 & 366684609  & M4V           & 10.92 & 3122.25      & 16.03         & 0.421             & 1.0                  & 12.64                         & 0.27                            & 23          & 4             \\
2MASS J21462206+3813047 & 407688400  & M5V           & 10.82 & 2971.27      & 7.04          & 0.214             & 1.4                  & 139.65                        & 0.07                            &             &               \\
2MASS J22232904+3227334 & 164170728  & M0V           & 10.37 & 3350.00       & 15.23         & 0.593             & 8.5                  & 20.02                         & 0.55                            & 83          & 4             \\
2MASS J23415498+4410407 & 2041185141 & M5V           & 10.37 & 2971.27      & 3.16          & 0.178             & 2.5                  & 49.98                         & 0.1                             & 18          &               \\
AD Leo                  & 95431305   & M3V           & 8.21  & 4363.00       & 4.97          & 0.422             & 3.5                  & 178.49                        & 0.55                            & 186         & 24            \\
AU Mic                  & 441420236  & M1V           & 7.84  & 3642.00       & 9.72          & 0.698             & 8.5                  & 72.64                         & 2.82                            & 327         & 16            \\
BD+33 1505              & 302820286  & M0V           & 9.35  & 3619.00       & 18.22         & 0.598             & 3.7                  & 84.92                         & 0.22                            & 1           &               \\
BD-02 2198              & 65673065   & M1V           & 9.12  & 3866.00       & 14.07         & 0.577             & 3.2                  & 69.81                         & 0.24                            & 61          & 10            \\
BX Cet                  & 278962913  & M2V           & 10.32 & 3275.05      & 7.22          & 0.279             & 3.0                  & 165.17                        & 0.02                            & 3           &               \\
CE Boo                  & 258105174  & M0V           & 9.13  & 3780.00       & 9.93          & 0.477             & 4.3                  & 23.71                         & 0.12                            & 48          &               \\
EE Leo                  & 393584005  & M4V           & 10.28 & 3122.25      & 6.97          & 0.293             & 2.6                  & 90.20                          & 0.24                            & 1           &               \\
EG Cam                  & 53186913   & M0V           & 9.41  & 3739.48      & 13.49         & 0.513             & 2.3                  & 104.56                        & 0.06                            & 2           &               \\
EQ Peg                  & 247985094  & M4V           & 9.04  & 3630.00       & 6.26          & 0.513             & 16.0                 & 98.97                         & 0.14                            & 731         & 5             \\
EV Lac                  & 154101678  & M4V           & 9.0   & 3122.25      & 5.05          & 0.337             & 5.1                  & 219.02                        & 0.23                            & 967         & 5             \\
G 168-31                & 236325118  & M3V           & 10.98 & 3429.20       & 36.91         & 0.655             & 1.1                  & 34.61                         & 0.11                            &             &               \\
G 214-14                & 119578924  & M0V           & 10.38 & 3739.48      & 23.71         & 0.513             & 1.7                  & 47.08                         & 0.22                            &             &               \\
G 234-57                & 341612752  & M1V           & 10.46 & 3429.20       & 21.05         & 0.4               & 2.0                  & 17.70                          & 0.14                            & 1           &               \\
G 32-5                  & 52005579   & M4V           & 11.4  & 3122.25      & 12.21         & 0.269             & 5.5                  & 158.46                        & 0.11                            &             &               \\
G 99-49                 & 282501711  & M3V           & 9.9   & 3275.05      & 5.21          & 0.261             & 5.7                  & 90.93                         & 0.87                            & 141         & 2             \\
GJ 1                    & 120461526  & M2V           & 7.68  & 3429.20       & 4.35          & 0.396             & 2.8                  & 110.92                        & 0.36                            & 5           & 5             \\
GJ 1074                 & 354430548  & M0V           & 10.15 & 3584.18      & 21.11         & 0.537             & 4.0                  & 46.89                         & 0.26                            &             &               \\
GJ 1105                 & 371272665  & M4V           & 10.67 & 3275.05      & 8.84          & 0.294             & 1.9                  & 59.04                         & 0.11                            & 3           &               \\
GJ 15 A                 & 440109725  & M2V           & 7.22  & 3605.50       & 3.56          & 0.406             & 3.7                  & 89.77                         & 0.66                            &             & 1             \\
GJ 176                  & 397354290  & M2V           & 9.0   & 3679.00       & 9.47          & 0.487             & 12.6                 & 175.00                         & 0.32                            & 11          & 1             \\
GJ 180                  & 246902275  & M2V           & 9.93  & 3275.05      & 11.94         & 0.413             & 1.7                  & 73.40                          & 0.55                            & 6           & 1             \\
GJ 184                  & 259737046  & M0V           & 9.21  & 3739.48      & 13.86         & 0.53              & 3.5                  & 69.73                         & 0.17                            & 3           &               \\
GJ 2                    & 439946126  & M2V           & 9.08  & 3875.00       & 11.5          & 0.515             & 1.8                  & 43.42                         & 0.14                            & 1           &               \\
GJ 205                  & 50726077   & M1V           & 7.1   & 3731.20       & 5.7           & 0.561             & 3.3                  & 44.54                         & 0.93                            & 8           &               \\
GJ 2066                 & 455139555  & M0V           & 9.12  & 3429.20       & 8.94          & 0.443             & 1.9                  & 22.01                         & 0.15                            &             &               \\
GJ 229                  & 124279525  & M1V           & 7.31  & 3814.00       & 5.76          & 0.549             & 3.1                  & 113.37                        & 2.34                            & 4           &               \\
GJ 26                   & 267688052  & M1V           & 10.05 & 3429.20       & 12.67         & 0.43              & 2.2                  & 43.27                         & 0.11                            &             &               \\
GJ 273                  & 318686860  & M4V           & 8.59  & 3275.05      & 3.79          & 0.316             & 2.2                  & 22.77                         & 0.66                            & 2           &               \\
GJ 317                  & 118608254  & M4V           & 10.75 & 3275.05      & 15.2          & 0.427             & 2.8                  & 171.04                        & 0.26                            & 5           &               \\
GJ 328                  & 265373654  & M0V           & 9.29  & 3739.48      & 20.54         & 0.651             & 3.4                  & 85.18                         & 0.34                            &             &               \\
GJ 3323                 & 43605290   & M4V           & 10.65 & 3122.25      & 5.38          & 0.186             & 2.3                  & 96.11                         & 0.52                            & 26          &               \\
GJ 358                  & 259999047  & M0V           & 9.63  & 3275.05      & 9.6           & 0.423             & 1.6                  & 208.15                        & 0.15                            & 160         &               \\
GJ 3649                 & 356668126  & M1V           & 9.88  & 3584.18      & 16.68         & 0.529             & 1.9                  & 40.88                         & 0.04                            &             &               \\
GJ 382                  & 77612635   & M0V           & 8.33  & 3429.20       & 7.7           & 0.51              & 2.2                  & 155.11                        & 1.41                            & 16          & 6             \\
GJ 3822                 & 72599316   & M1V           & 9.83  & 3584.18      & 20.34         & 0.581             & 3.5                  & 34.44                         & 0.12                            &             &               \\
GJ 399                  & 54959262   & M1V           & 10.26 & 3429.20       & 15.58         & 0.466             & 1.7                  & 132.57                        & 0.4                             & 2           &               \\
GJ 3997                 & 230231375  & M1V           & 9.64  & 3739.48      & 13.63         & 0.48              & 2.7                  & 77.36                         & 0.63                            &             &               \\
GJ 408                  & 97472519   & M2V           & 8.97  & 3122.25      & 6.75          & 0.39              & 2.1                  & 22.40                          & 0.26                            & 1           &               \\
GJ 4092                 & 222495959  & M0V           & 10.12 & 3739.48      & 28.23         & 0.63              & 2.7                  & 24.49                         & 0.69                            &             &               \\
GJ 422                  & 450545735  & M4V           & 10.48 & 3275.05      & 12.67         & 0.37              & 1.2                  & 175.46                        & 0.13                            & 1           &               \\
GJ 433                  & 57654763   & M0V           & 8.89  & 3616.00       & 9.07          & 0.469             & 1.3                  & 170.30                         & 0.64                            &             &               \\
GJ 436                  & 138819293  & M1V           & 9.57  & 3416.00       & 9.76          & 0.425             & 1.7                  & 60.04                         & 0.07                            & 1           &               \\
GJ 450                  & 144400022  & M1V           & 8.85  & 3584.18      & 8.76          & 0.46              & 5.8                  & 60.77                         & 0.86                            & 21          & 1             \\
GJ 47                   & 256316623  & M2V           & 9.84  & 4104.00       & 10.52         & 0.39              & 2.0                  & 71.04                         & 0.06                            & 1           &               \\
GJ 49                   & 256419669  & M2V           & 8.66  & 4055.50       & 9.86          & 0.54              & 2.9                  & 172.89                        & 0.39                            & 31          &               \\
GJ 494                  & 88138162   & M0V           & 8.91  & 3899.50       & 11.51         & 0.563             & 9.1                  & 50.20                          & 0.7                             & 165         & 7             \\
GJ 514                  & 404519959  & M1V           & 8.21  & 3727.00       & 7.62          & 0.503             & 1.9                  & 34.48                         & 0.86                            & 3           & 1             \\
GJ 521                  & 458452740  & M2V           & 9.4   & 3584.18      & 13.37         & 0.498             & 2.9                  & 81.41                         & 0.05                            &             &               \\
GJ 526                  & 72465347   & M2V           & 7.61  & 3634.00       & 5.44          & 0.482             & 2.4                  & 19.48                         & 0.68                            & 2           &               \\
GJ 536                  & 119147875  & M0V           & 8.86  & 4067.00       & 10.41         & 0.508             & 1.7                  & 65.31                         & 0.4                             & 1           & 2             \\
GJ 552                  & 450352108  & M1V           & 9.72  & 3429.20       & 14.25         & 0.503             & 2.6                  & 30.79                         & 0.14                            &             &               \\
GJ 581                  & 36853511   & M1V           & 9.41  & 3442.00       & 6.3           & 0.33              & 1.8                  & 54.78                         & 0.43                            & 4           &               \\
GJ 588                  & 149590989  & M3V           & 8.27  & 3429.20       & 5.92          & 0.46              & 1.8                  & 97.28                         & 0.47                            & 9           &               \\
GJ 606                  & 49511355   & M0V           & 9.59  & 3584.18      & 13.29         & 0.487             & 2.0                  & 35.19                         & 0.46                            & 1           & 5             \\
GJ 628                  & 413948621  & M3V           & 8.79  & 3570.00       & 4.31          & 0.322             & 1.5                  & 35.44                         & 0.62                            & 4           & 2             \\
GJ 649                  & 236743275  & M2V           & 8.82  & 3696.33      & 10.38         & 0.517             & 2.1                  & 96.02                         & 0.52                            & 13          &               \\
GJ 65                   & 632499596  & M5V           & 10.81 & 2971.27      & 2.72          & 0.165             & 26.4                 & 44.81                         & 1.27                            & 153         & 35            \\
GJ 674                  & 218263393  & M3V           & 8.33  & 3275.05      & 4.55          & 0.365             & 1.8                  & 117.30                         & 0.46                            & 82          & 1             \\
GJ 676 A                & 369021303  & M0V           & 8.87  & 3739.48      & 16.03         & 0.649             & 2.6                  & 115.98                        & 0.3                             & 6           & 2             \\
GJ 686                  & 230241767  & M1V           & 8.74  & 3584.18      & 8.16          & 0.442             & 2.9                  & 99.75                         & 1.04                            & 1           &               \\
GJ 699                  & 325554331  & M1V           & 8.2   & 3244.67      & 1.83          & 0.194             & 2.5                  & 37.89                         & 2.06                            & 3           & 5             \\
GJ 70                   & 369534824  & M1V           & 9.9   & 3429.20       & 11.32         & 0.408             & 2.0                  & 37.73                         & 0.21                            &             &               \\
GJ 701                  & 41378014   & M0V           & 8.52  & 3630.00       & 7.73          & 0.465             & 1.9                  & 36.60                          & 1.79                            & 4           & 1             \\
GJ 731                  & 298514156  & M0V           & 9.38  & 3739.48      & 15.21         & 0.539             & 2.7                  & 97.78                         & 0.72                            &             &               \\
GJ 740                  & 227642705  & M1V           & 8.46  & 3584.18      & 11.11         & 0.588             & 2.3                  & 50.94                         & 1.31                            & 1           &               \\
GJ 752 A                & 176188045  & M3V           & 8.1   & 3275.05      & 5.91          & 0.473             & 2.7                  & 96.89                         & 1.51                            & 6           &               \\
GJ 83.1                 & 404715018  & M5V           & 10.67 & 3122.25      & 4.47          & 0.18              & 2.6                  & 42.63                         & 0.35                            & 18          & 1             \\
GJ 832                  & 139754153  & M2V           & 7.74  & 3707.00       & 4.97          & 0.442             & 2.0                  & 112.46                        & 0.55                            & 5           & 1             \\
GJ 846                  & 292987389  & M0V           & 8.4   & 3580.00       & 10.55         & 0.574             & 3.1                  & 92.18                         & 1.79                            &             &               \\
GJ 849                  & 248027247  & M0V           & 9.22  & 3275.05      & 8.8           & 0.464             & 1.7                  & 116.90                         & 0.72                            & 5           &               \\
GJ 876                  & 188580272  & M3V           & 8.88  & 3532.00       & 4.68          & 0.352             & 2.5                  & 180.85                        & 0.83                            & 3           &               \\
GJ 880                  & 217331269  & M1V           & 7.79  & 3750.00       & 6.87          & 0.55              & 2.4                  & 53.34                         & 0.59                            & 7           & 3             \\
GJ 908                  & 267255253  & M1V           & 8.15  & 3646.00       & 5.9           & 0.417             & 2.6                  & 72.40                          & 0.72                            & 5           &               \\
GJ 9122 B               & 434136638  & M0V           & 9.92  & 3739.48      & 17.24         & 0.523             & 3.6                  & 103.32                        & 0.06                            &             &               \\
GJ 9404                 & 328860504  & M0V           & 9.87  & 3739.48      & 23.9          & 0.626             & 2.6                  & 60.13                         & 0.03                            & 2           & 1             \\
GJ 9793                 & 467549713  & M0V           & 10.04 & 3739.48      & 31.4          & 0.692             & 1.0                  & 26.10                          & 0.45                            & 3           & 3             \\
Gl 799B                 & 1992325967 & M4V           & 9.59  & 3123.00       & 9.83          & 0.692             & 10.2                 & 92.19                         & 0.11                            & 535         &               \\
Gl 841 A                & 140045538  & M2V           & 9.4   & 3429.20       & 14.86         & 0.608             & 4.2                  & 110.37                        & 0.45                            & 433         & 15            \\
HD 154363B              & 142494324  & M1V           & 9.17  & 3584.18      & 10.46         & 0.463             & 2.7                  &                               & 0.5                             &             & 9             \\
HD 233153               & 311064351  & M1V           & 8.91  & 5125.96      & 12.28         & 0.555             & 2.7                  & 81.19                         & 1.1                             & 3           & 39            \\
HD 265866               & 68581262   & M1V           & 8.86  & 3275.05      & 5.58          & 0.368             & 1.7                  & 188.22                        & 0.67                            & 4           & 2             \\
HD 50281B               & 282210724  & M0V           & 9.09  & 4763.86      & 8.74          & 0.442             & 3.9                  &                               & 0.21                            &             & 12            \\
HD 79211                & 251078597  & M0V           & 7.05  & 3995.02      & 6.33          & 0.586             &                      & 49.21                         & 0.21                            &             &               \\
HD 95735                & 166646191  & M2V           & 6.55  & 3563.50       & 2.55          & 0.389             & 7.3                  & 43.97                         & 0.56                            &             &               \\
HIP 57050               & 115869504  & M4V           & 10.58 & 3122.25      & 11.02         & 0.359             & 1.8                  & 59.62                         & 0.11                            &             &               \\
HIP 79431               & 49324530   & M1V           & 10.24 & 3275.05      & 14.54         & 0.479             & 1.0                  & 35.42                         & 0.28                            &             &               \\
LHS 3432                & 384491142  & M0V           & 9.8   & 3429.20       & 8.82          & 0.336             & 4.3                  & 70.04                         & 0.35                            & 2           &               \\
LP 609-71               & 841962647  & M1V           & 9.61  & 3429.20       & 11.54         & 0.485             & 2.7                  & 53.43                         & 0.24                            &             & 3             \\
LP 672-42               & 285972083  & M3V           & 10.81 & 3275.05      & 13.44         & 0.372             & 1.5                  & 163.44                        & 0.06                            & 7           &               \\
MCC 549                 & 17122372   & M0V           & 10.28 & 3739.48      & 38.8          & 0.815             & 19.1                 & 45.89                         & 0.32                            & 4           &               \\
Proxima Cen        & 1019422535 & M4V           & 8.95  & 2990.50       & 1.3           & 0.154             & 2.6                  & 153.35                        & 0.24                            & 364         &               \\
Ross 733                & 353011288  & M4V           & 10.37 & 3122.25      & 18.1          & 0.519             & 14.0                 & 112.02                        & 0.13                            & 252         &               \\
TYC 1313-1482-1         & 717082161  & M0V           & 10.27 & 3739.48      & 46.23         & 0.870             & 1.0                  & 20.56                         & 0.14                            & 7           &               \\
TYC 4902-210-1          & 33168946   & M0V           & 10.01 & 3739.48      & 30.67         & 0.706             & 1.6                  & 40.76                         & 0.27                            & 25          &               \\
V 1054 Oph              & 181589861  & M3V           & 7.91  & 3200.00       & 6.2           & 0.533             & 2.1                  &                               & 0.64                            &             & 22            \\
V 1352 Ori              & 247463344  & M4V           & 10.1  & 3122.25      & 5.79          & 0.249             & 4.7                  & 196.00                         & 0.11                            & 1           &               \\
VV Lyn                  & 16034014   & M2V           & 10.47 & 3429.20       & 11.87         & 0.518             & 4.6                  & 87.78                         & 0.45                            & 58          &               \\
Wolf 906                & 35400809   & M1V           & 10.17 & 3429.20       & 14.46         & 0.461             & 1.7                  & 86.00                          & 0.22                            & 1           &               \\
YZ Ceti                 & 610210976  & M5V           & 10.43 & 3122.25      & 3.71          & 0.168             & 2.2                  & 44.98                         & 1.22                            & 54          & 10     

\label{annex:table}
\end{longtable}
\end{landscape}

\section{Additional complex flare decompositions}
\begin{figure}[h]
\centering
\includegraphics{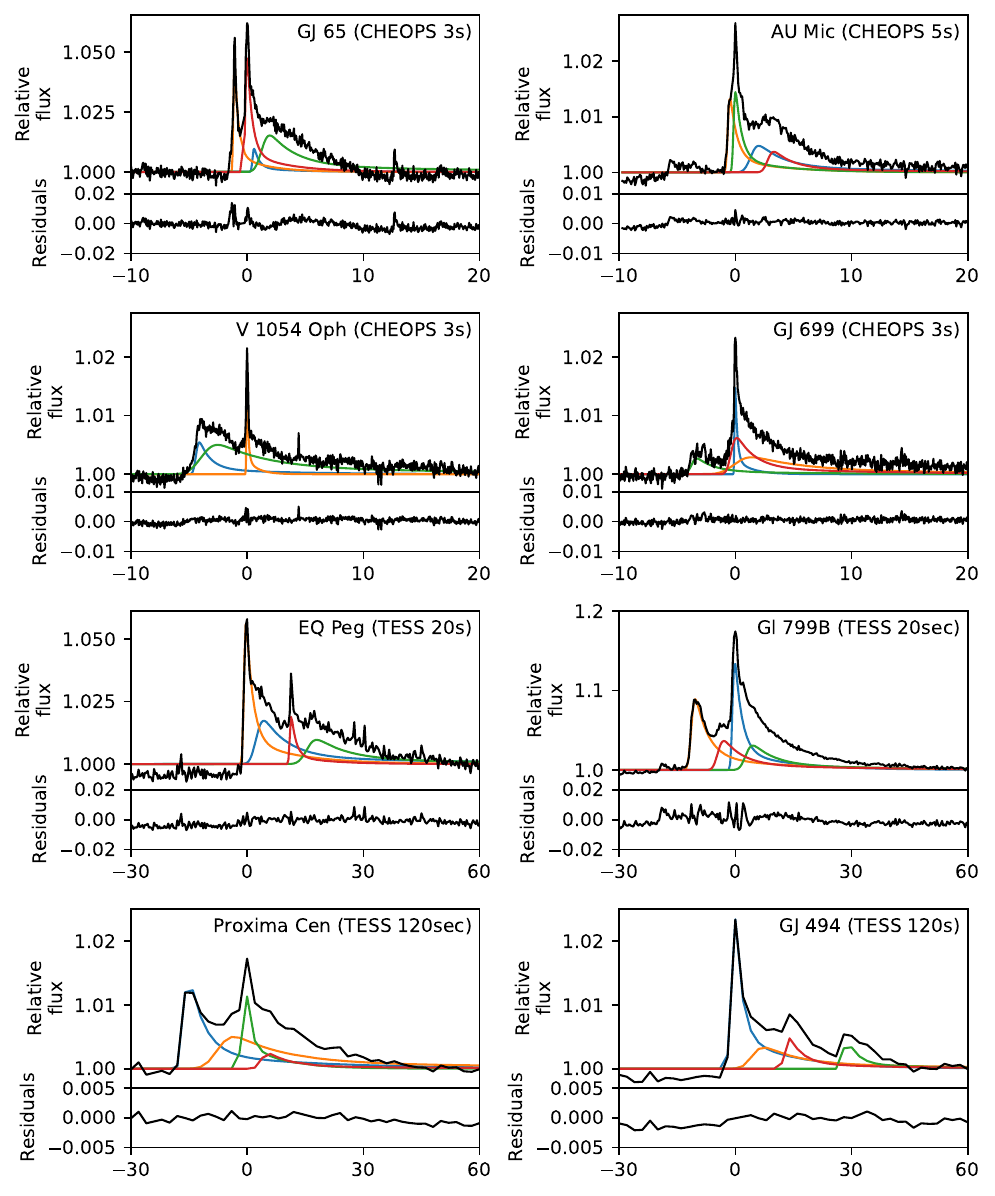}
\caption{Additional fits of complex flares. The panel titles indicate the star, the detecting mission, and the observational cadence for each event. The observed light curve is plotted in black, with the fitted flare components overplotted in colour. The colours carry no physical meaning and simply indicate the order in which components were added during the fitting, which depends on the specific flare shape. The residuals corresponding to each fit are displayed below the light curves.}
\label{fig:multiflare}
\end{figure}

\clearpage
\section{FFDs, fits and tests for the most active subsample}
\begin{figure*}[h]
\centering
\includegraphics[width=0.95\textwidth]{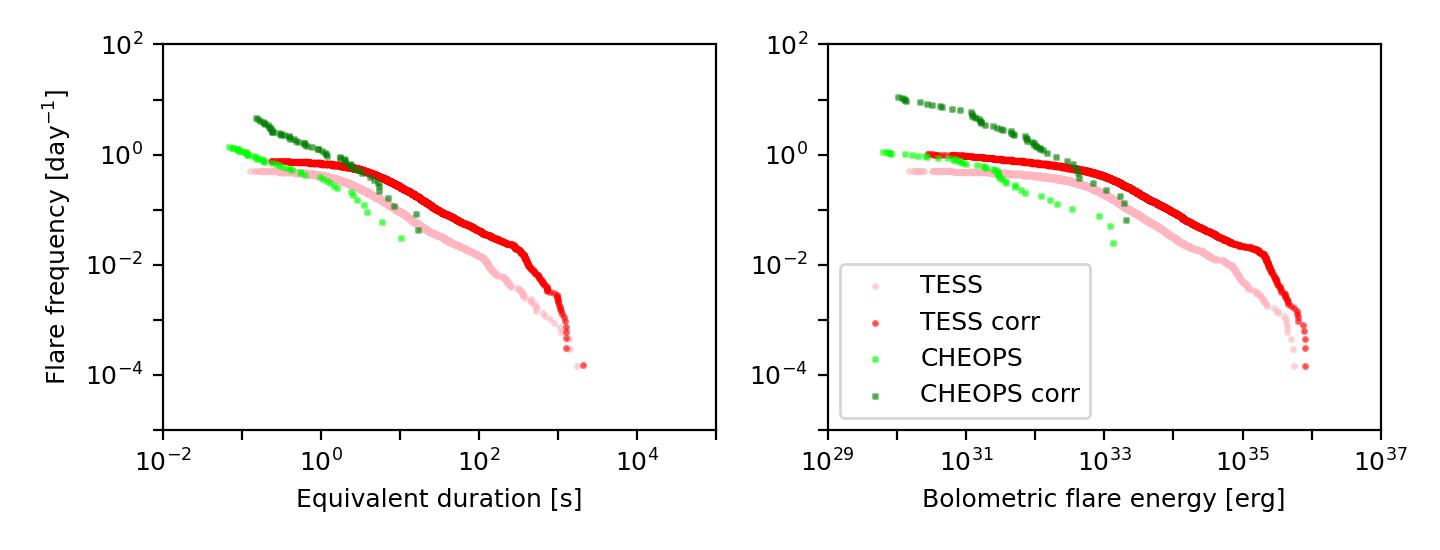}
\caption{Same as Figure \ref{fig:ffds}, but for the active subsample.}
\label{fig:ffds_subsample}
\end{figure*}

\begin{figure*}[h]
\centering
\includegraphics[width=0.95\textwidth]{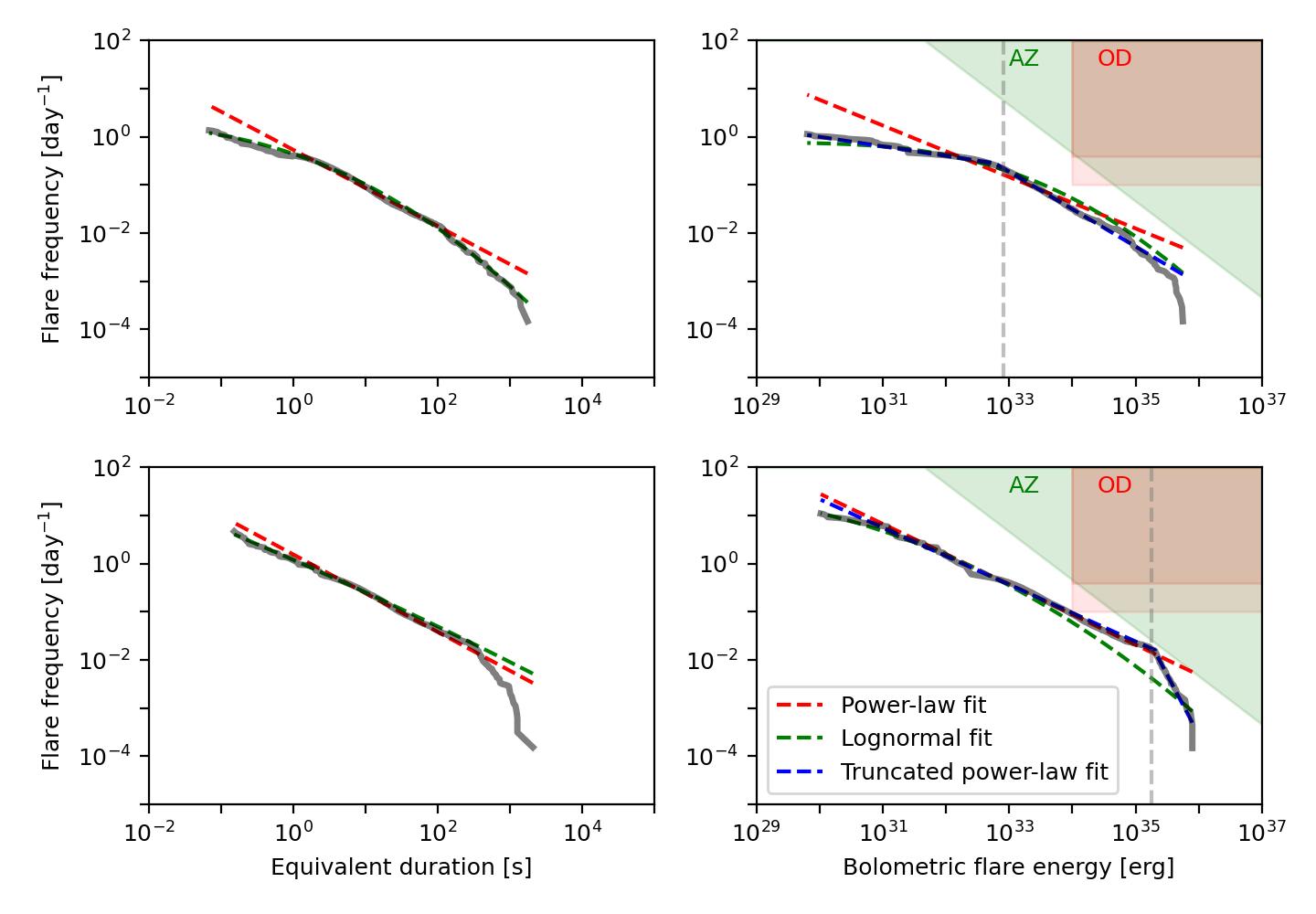}
\caption{Same as Figure \ref{fig:fits}, but for the active subsample. We note a deviation from a power law at high EDs in both the observed (top-left) and corrected (bottom-left) distributions, which was not visible considering the full sample. We attribute this deviation to a sampling bias that becomes noticeable at high EDs when the number of targets is reduced. The uncorrected observational bias at low EDs in the observed distribution leads to a preference for a lognormal fit over a power law, representing the only difference from the conclusions drawn using the full sample. The obtained truncation thresholds obtained are $t_1=6.0\times10^{32}$ and $t_2=2.0\times10^{35}$ erg.}
\label{fig:fits_subsample}
\end{figure*}
\FloatBarrier

\clearpage
\begin{table*}[h]
\caption{Same as Table \ref{table:fits}, but restricted to the active subsample.}
\begin{adjustbox}{width=\textwidth}
\begin{tabular}{lccccccccc}
\hline
     FFD                       & Power law             & Lognormal            & Truncated            & R$_{PL}$               & p$_{PL}$          & R$_{LT}$               & p$_{LT}$    & $\mathrm{R}_{PT}$               & $\mathrm{p}_{PT}$       \\ \hline
TESS (ED)                   &      $\alpha=1.70$                &       $\mu=0.45, \ \sigma=0.57$               &     --                 &       -8.40               &      < 0.001                &       --               &         --             \\
TESS corr. (ED)         &     $\alpha=1.67$                 &       $\mu=0.77, \ \sigma=0.59$               &     --                 &      -11.59                &     < 0.001                &        --              &          --            \\
CHEOPS (ED)                 &    $\alpha=1.63$                  &       $\mu=-18.60, \ \sigma=3.86$               &    --                  &         -0.72             &      0.047             &       --               &         --             \\
CHEOPS corr. (ED)       &    $\alpha=1.80$                  &       $\mu=-15.85, \ \sigma=3.09$             &   --                   &       -0.14              &      0.089               &       --               &         --             \\
TESS (bol. E)               &      $\alpha=1.47$                &      $\mu=32.82, \ \sigma=0.71$                &             --         &         -2.00           &     0.126                   &        --              &         --             \\
TESS corr. (bol. E)     &     $\alpha=1.48$                 &      $\mu=32.84, \ \sigma=0.93$                &             --         &     10.03                &       0.041                &         --             &             --         \\
CHEOPS (bol. E)             &      $\alpha=1.43$                &   $\mu=31.27, \ \sigma=0.76$                   &              --        &       -0.95               &     0.089                   &    --                  &        --              \\
CHEOPS corr. (bol. E)   &       $\alpha=1.58$               &      $\mu=30.67, \ \sigma=1.05$                  &            --          &     2.76                  &     0.099                  &      --                &       --               \\
Combined (ED)               & $\alpha=1.79$ & $\mu=-0.91, \ \sigma=1.79$ & -- & 7.16 & 0.045 & -- & -- \\
Combined corr. (ED)     & $\alpha=1.80$ & $\mu=-19.33, \ \sigma=3.63$ & -- & -6.40 & 0.078 & -- & -- \\
Combined (bol. E)           & $\alpha=1.53$ & $\mu=32.14, \ \sigma=1.25$ & $\alpha_{<t_1}=1.20, \ \alpha_{>t_1}=1.78$ & 2.40 & 0.016 & 1.44 & 0.015 & 3.85 & 0.012 \\
Combined corr. (bol. E) & $\alpha=1.63$ & $\mu=28.91, \ \sigma=1.70$ & $\alpha_{<t_2}=1.59, \ \alpha_{>t_2}=3.49$ & -0.77 & 0.443 & 0.99 & 0.032 & 0.22 & 0.042 \\ \hline
\label{table:fits_subsample}
\end{tabular}
\end{adjustbox}
\end{table*} 

\begin{figure*}[h]
\centering
\includegraphics{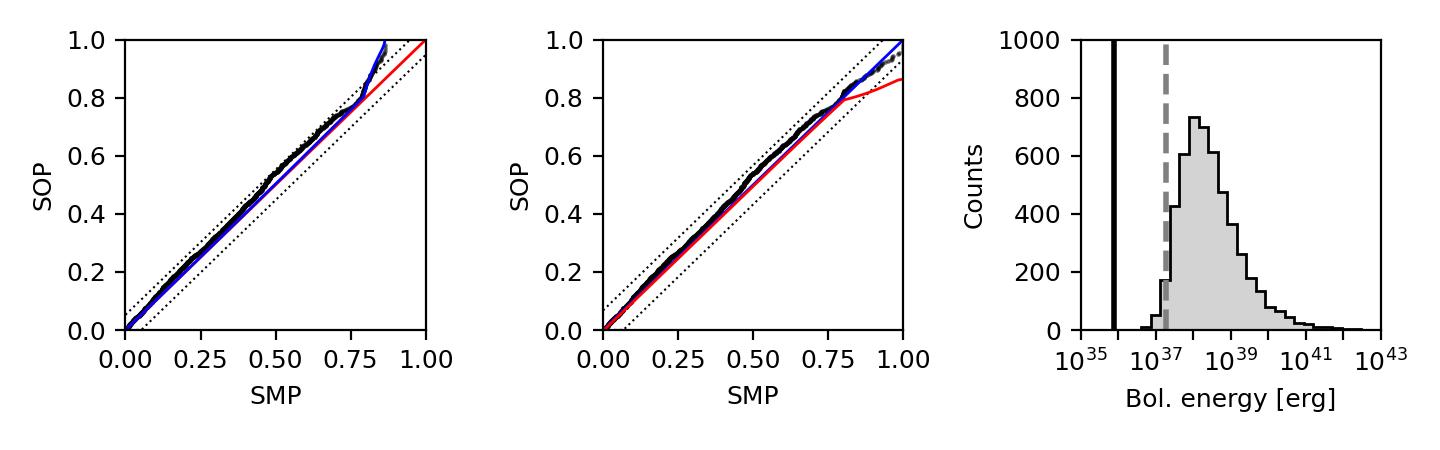}
\caption{Same as Figures \ref{fig:sks}, but for the active subsample. In the left and middle panels, red and blue lines correspond to the best-fit infinite and truncated power laws, respectively. We observe a deviation from the infinite power law at high percentiles, which the exceedance test indicates is caused by limited sampling.}
\label{fig:tests_subsample}
\end{figure*}

\end{document}